\documentclass[preprint,3p,times]{elsarticle}
\usepackage[utf8]{inputenc}
\usepackage{CJKutf8}
\usepackage{amssymb}
\usepackage{amsmath,amssymb}
\usepackage{algpseudocode}
\usepackage{algorithm}
\usepackage[hyphens]{url}
\usepackage{graphicx}
\usepackage{comment}
\usepackage{bm}
\newcommand{\vect}[1]{\boldsymbol{\mathbf{#1}}}
\def\vec#1{\vect{#1}}
\usepackage{lineno}
\usepackage{color}
\newcommand\HL[1]{{\color{black}#1}}

\journal{journal}

\begin{document}

\begin{frontmatter}

%\title{A hydrodynamic approach to numerically reproduce multiple spinning vortices in horizontally rotating three-dimensional liquid helium-4}
\title{A hydrodynamic approach to reproduce multiple spinning vortices in horizontally rotating three-dimensional liquid helium-4}
	
\author[RCAST]{Satori Tsuzuki~(\begin{CJK}{UTF8}{min}都築怜理\end{CJK})}
\ead{tsuzukisatori@g.ecc.u-tokyo.ac.jp}

\address[RCAST]{Research Center for Advanced Science and Technology, The University of Tokyo, 4-6-1, Komaba, Meguro-ku, Tokyo 153-8904, Japan}

\begin{abstract}
This paper reports a three-dimensional (3D) simulation of a rotating liquid helium-4, using a two-fluid model with spin-angular momentum conservation. Our model was derived from the particle approximation of an inviscid fluid with residual viscosity. 
Despite the fully classical mechanical picture, the resulting system equations were consistent with those of the conventional two-fluid model. 
We consider bulk liquid helium-4 to be an inviscid fluid, assuming that the viscous fluid component remains at finite temperatures. As the temperature decreased, the amount of the viscous fluid component decreased, ultimately becoming a fully inviscid fluid at absolute zero. 
Weak compressibility is assumed to express the volume change because some helium atoms do not render fluid owing to Bose-Einstein condensations (BECs) or change states because of local thermal excitation. One can solve the governing equations for an incompressible fluid using explicit SPH (smoothed-particle hydrodynamics), simultaneously reproducing density fluctuations and describing the fluid in a many-particle system. We assume the following fluid-particle duality: a hydrodynamic interfacial tension between the inviscid and viscous components or a local interaction force between two types of fluid particles. The former can be induced in the horizontal direction when non-negligible non-uniformity of the particles occurs during forced two-dimensional rotation, and the latter is non-negligible when the former is negligible. We performed a large-scale simulation of 3D liquid helium forced to rotate horizontally using 32 number of graphics processing units (GPUs). Compared with the low-resolution calculation using 2.4 million particles, the high-resolution calculation using 19.6 million particles showed spinning vortices close to those of the theoretical solution. We obtained a promising venue to establish a practical simulation method for bulk liquid helium-4.
\end{abstract}


\end{frontmatter}

%\linenumbers

%\section{}
%\label{}





\section{Introduction}
Liquid helium-4 has a long history of use in various engineering and scientific fields. In particular, it has been used as a refrigerant for superconducting magnets in magnetic resonance imaging (MRI) and has been studied for cooling superconducting motors for rapid long-distance mobility. In general, two types of cooling methods are used in superconducting systems. The first is low-temperature superconductivity, an example of which is a niobium-titanium alloy cooled by liquid helium to near absolute zero (0 to about 4.2 K: Kelvin)~\cite{Arzhavitin2016, BANNO2023100047}. The second type is high-temperature superconductivity, in which a special metallic compound is cooled to a relatively high temperature of approximately 100 K using liquid nitrogen to achieve a superconducting state. Low-temperature superconductivity has the advantage of high critical current density. However, according to quantum mechanics, liquid helium loses its viscosity when cooled to near absolute zero owing to Bose-Einstein condensation to the ground state of energy, and this quantum effect often causes a severe leakage phenomenon (super leakage). In addition, it only enters the solid state under high pressure and thus must be used as a coolant in the liquid state. In other words, the stable control of liquid helium in a cooling system is very difficult in low-temperature superconductivity. Stable cooling with liquid nitrogen enables high-temperature superconductivity. However, the critical current density is small, and the current that can maintain the superconducting state is weak; therefore, it cannot withstand practical use. Thus, neither cooling method is one step away from practical application owing to the trade-off between the difficulty of cooling and the critical current density; this is a factor that hinders the full industrialization of superconducting motors.

If it were possible to perform large-scale simulations on recent supercomputers to reproduce the dynamics of bulk liquid helium-4, the following would be expected. First, it would be possible to precisely predict and control the occurrence of super leakage due to quantum effects in the dynamics of superfluid helium; this will enable the stable control and safety of superfluid helium in large-scale cooling systems at a low development cost. Consequently, the practicality of superfluid helium as a cold agent in low-temperature superconductivity is expected to be dramatically enhanced. In addition, it is expected to significantly reduce the consumption of liquid helium during cryogenic experiments. It has been reported that the carbon dioxide emitted from the production and transportation of liquid helium can reach 587 kg-CO2/l~\cite{10.1063/10.0020164}. If the dynamics of bulk liquid helium can be simulated by large-scale simulation technology using a supercomputer, the number of experiments using liquid helium can be reduced by half, leading to a reduction in carbon dioxide emissions.

This study developed a new simulation technique that can reproduce the three-dimensional (3D) fluid behavior of superfluid helium-4 in a cryogenic cooling system on a real scale. As of yet, most simulation methods for liquid helium have been based on quantum mechanical scale simulations that aim to elucidate the principles, and there has been little progress in the development of numerical methods for so-called ``bulk state'' quantum fluids. Next, we discuss the properties of helium-4 at cryogenic temperatures. In principle, the mechanism of liquid helium can be described using quantum mechanics, which is the governing law of the nanoscale. Helium-4 is a so-called Bose particle consisting of two neutrons and two protons. Bose particles have quantized angular momentum times an integer. In this paper, ``helium'' indicates helium-4 unless otherwise noted. Liquid helium-4 can be described as a many-particle interaction system of Bose particles. However, unlike ordinary materials that solidify in low-temperature regions, helium-4 exists as a liquid even near absolute zero. Therefore, liquid helium has the strange property of locally obeying quantum mechanical laws while behaving as a fluid, a continuum, as a whole. In a many-particle interacting system of Bosonic particles below a critical temperature near absolute zero, helium atoms in a particular region undergo Bose-Einstein condensation (BEC), in which they fall into the ground energy state; this minimizes the interaction between helium atoms. (Helium is a monatomic molecule, but we use the term ``molecular viscosity'' in accordance with the chemical engineering doctrine~\cite{WilsonLloyd1950, Immanuil-L-Fabelinskii_1997}). In other words, liquid helium-4 loses its viscosity at cryogenic temperatures below the critical temperature, that is, it becomes an inviscid fluid.

The fact that liquid helium becomes an inviscid fluid causes peculiar fluid phenomena, such as slipping through the microstructure of the porous medium (gypsum), which is not observed when liquid helium is composed only of viscous normal flow components, and film flow caused by the viscosity between the wall surface and helium atoms being larger than the viscosity between helium atoms. These phenomena are often described as ``macroscopic quantum phenomena'' because the disappearance of microscopic interactions, i.e., of molecular viscosity, affects the macroscopic fluid behavior of bulk liquid helium. In addition to the phenomena that can be observed with the naked eye, vortex lattice phenomena observed on the nanometer scale in rotating liquid helium are also macroscopic quantum phenomena that can be observed between the hydrodynamic and quantum mechanical regimes. To the best of our knowledge, no study has successfully simulated the vortex lattice phenomena of rotating 3D liquid helium-4 observed in the boundary regime between these two different dynamical regimes using a method based on classical fluid mechanics.

It is widely recognized that the macroscopic behavior of liquid helium-4 can be phenomenologically described by the two-fluid model proposed by L. Landau, which consists of an inviscid fluid and incompressible Navier-Stokes equations. However, it should be noted that this assumption has only been verified by the issue of countercurrent flow, in which shear viscosity is dominant. Macroscopic quantum phenomena observed in the rotation problem, such as vortex lattices, have yet to be numerically reproduced using two-fluid models. The authors hypothesized that the main reason for this is that the ordinary two-fluid model does not reflect the quantization of the spin angular momentum of individual helium-4 atoms when liquid helium-4 is considered as a microscopic many-particle interacting system. As stated previously, the spin angular momentum of the Bose particles was quantized. If fluid particles, which are virtual constituent particles in fluid mechanics, can be interpreted as a type of coarse-grained model of such microscopic particles, then the angular momentum related to the spin of the constituent atoms should also be conserved in the fluid particles.

Considering this hypothesis, the author recently developed a two-fluid model with spin-angular momentum conservation~\cite{Tsuzuki_2021, doi:10.1063/5.0060605, doi:10.1063/5.0122247}. The following overview provides a general introduction to this subject. The Navier-Stokes equation with conservation of rotational angular momentum around the axes of fluid particles was originally derived from the constitutive law of continuum mechanics by D. W. Condiff in 1964~\cite{doi:10.1063/1.1711295}. Half a century later, K. M{\"{u}}ller~\cite{MULLER2015301} developed a particle approximation of the Navier-Stokes equation with spin-angular momentum conservation using smoothed-particle hydrodynamics (SPH)~\cite{gingold1977smoothed, monaghan1992smoothed} for mesoscale thermal flow. Considering these studies, the author adopted the Navier-Stokes equations with spin-angular momentum conservation for normal fluids and employed SPH to discretize the equation systems using particles. In other words, we describe liquid helium-4 as a system of numerous particle interactions composed of two distinct types of classical fluid particles that maintain their rotational angular velocity around their axes under the assumption of rigid body rotations; this is referred to as a two-fluid model with spin-angular momentum conservation. More importantly, our two-fluid model differs from conventional models in terms of the physical picture. The conventional two-fluid model was derived by combining the nonlinear $\rm Schr\ddot{o}dinger$ equation for bosons with the thermodynamic Gibbs--Duhem equation. Conversely, our model considers the bulk of liquid helium-4 at low temperatures as an inviscid fluid following continuum mechanics, which contains residual viscosity at finite temperatures. In summary, the proposed model is based on a two-phase flow. Despite the fully classical mechanical picture, the resulting system equations were consistent with an ordinary model (see Eqs.~(\ref{eq:goveqsuper:mut}) and (\ref{eq:goveqnormal:mut})). In particular, we consider the bulk liquid helium-4 to be an inviscid fluid, assuming that the viscous fluid component remains at finite temperatures. As the temperature decreased, the amount of the viscous fluid component decreased, ultimately becoming a fully inviscid fluid at absolute zero. From the Lagrangian perspective, our two-fluid model is a mixture of fluid particles from inviscid and viscous fluid components. Both fluid particles are classical fluid particles; they are virtual particles. Thus, our model is based on the mixing of two classical fluids and is similar to a two-phase flow model. However, as described below, we allow for a certain ``nature of particles'' of both components as a correction. In this context, the physical picture in which the viscous fluid component remains bead-like in an inviscid fluid is a more accurate representation of the system than a two-phase flow model. This is also close to the image of a ``nanofluid,'' or nanoparticle dispersion system, \HL{where the nanoparticles maintain fluidity.} Details of this technique are described in Refs.~\cite{Tsuzuki_2021, doi:10.1063/5.0060605, doi:10.1063/5.0122247}. This will be outlined later in this paper. In our previous simulations of rotating two-dimensional (2D) liquid helium-4, we observed the formation of multiple spinning vortices that occurred independently of the overall fluid motion. After incorporating the vortex dynamics model into our model, we found that multiple spinning vortices formed a lattice with a regular arrangement. The number of vortices generated is determined by the strength of the rotational angular velocity. For appropriate values, the number of vortices agrees well with the theoretical solution of the number of vortices calculated from Feynman's law. It is also noteworthy that the spinning of each vortex moves independently without affecting the behavior of the entire fluid. In other words, the vortices do not dissipate. In a quantum vortex, the vortices do not dissipate because of the circulation quantization. This important property can also be reproduced. Consequently, the vortex lattice phenomenon in the rotating 2D liquid helium-4 was simulated by solving the two-fluid model with spin angular momentum conservation.

This study presents the observation of independently spinning three-dimensional vortices in 3D liquid helium-4 in a cylindrical container rotated horizontally using the aforementioned two-fluid model with spin angular momentum conservation. In the context of quantum hydrodynamics, the dynamics of the connections or reconnections of 3D vortex lines differ from those in classical dynamics owing to quantized circulations. Consequently, the topological variations in the 2D vortex lines are not analogous to those in the 3D vortex lines. For example, two vortex lines can occupy skew positions relative to one another in space, resulting in only two distinct patterns of reconnection between vortex line-line interactions and vortex line-wall interactions. In addition, from the perspective of polyhedral geometry, angle management in 3D space requires the use of quaternions to circumvent singularities. The dynamics of quantum fluids in 3D space are complex and, in fact, not an extension of those in 2D space; the cross-section of a vortex line in real cases is projected onto a plate in 2D space. The geometric complexity in 3D space indicates that computational methods in 3D space must be verified and validated independently of the results in 2D cases. Therefore, it is unclear whether the authors' method, which was effective in solving the 2D rotating problem, can numerically reproduce the dynamics of 3D liquid helium-4. For clarity, we excluded the vortex dynamics model effective for 2D liquid helium-4 designed in our previous studies~\cite{doi:10.1063/5.0060605} when applying our two-fluid model to 3D cases. We only focused on discretizing our two-fluid model in 3D space. Artificial manipulations such as those proposed by K. W. Schwarz~\cite{PhysRevB.31.5782, PhysRevB.38.2398} were not considered in this study. Nevertheless, we succeeded in reproducing several 3D vortices spinning in the horizontal direction during the horizontal rotation of liquid helium-4. The scientific significance of our numerical results is discussed in Section~\ref{sec:Discuss}. The remainder of this paper is organized as follows. In Section 2, we provide an overview of the proposed two-fluid model and the 3D calculations on a parallel computing platform. In Section 3, we present the numerical results of our simulations of 3D liquid helium-4 rotating horizontally. In Section 4, we discuss the scientific significance of the results obtained. Finally, Section 5 concludes the study.

\section{Methods} \label{seq:methods}
\subsection{The two-fluid model with spin angular momentum conservation}
\begin{figure}[t]
\vspace{-56.5cm}
\hspace{28.8cm}
%\begin{center}
\centerline{\includegraphics[width=4.5\textwidth, clip, bb= 0 0 4380 4048]{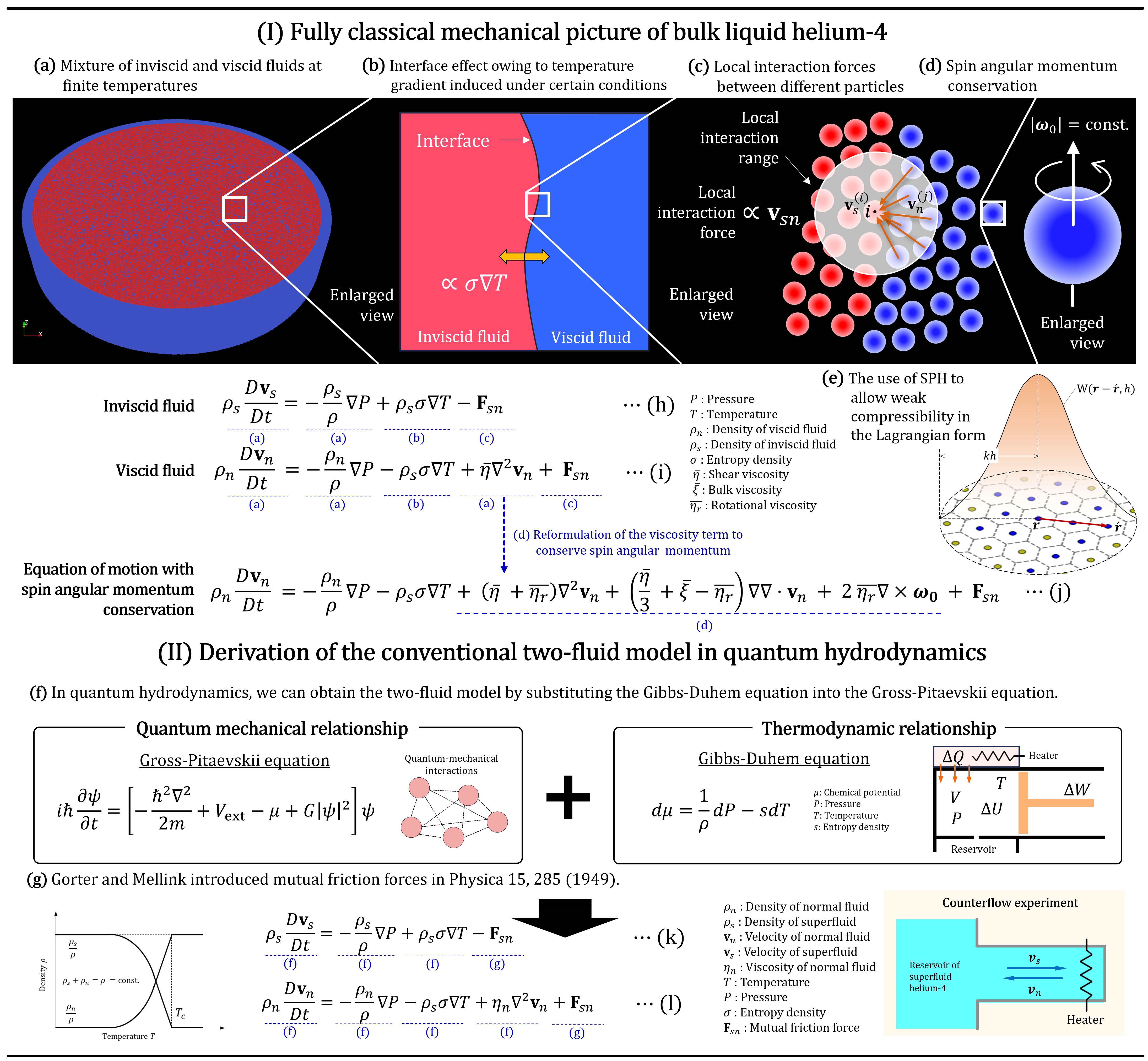}}
\caption{\HL{Schematic of our methodology: (I) a classical mechanical interpretation of bulk liquid helium-4, and (II) the derivation of the conventional two-fluid model in quantum hydrodynamics for comparison. Each part presents schematics of (a) the mixture of inviscid and viscid fluids, (b) the interfacial effect due to the temperature gradient induced under certain conditions, (c) the local interaction forces between different types of fluid particles, (d) the spin-angular momentum conservation, and (e) the use of the explicit SPH to realize the density fluctuation (weak compressibility). Next, (f) visually illustrates the derivation of the original two-fluid model from the Gross--Pitaevskii equation and the Gibbs--Duhem thermodynamic relation, (g) describes the introduction of mutual frictional forces. (h) is the Euler equation for an inviscid fluid, (i) is the Navier-Stokes equation for a viscous fluid, and (j) is a reformulated version of (i) introducing spin-angular momentum conservation; (h)-(j) are derived from the classical mechanical assumption. In contrast, (k) and (l) represent the ordinary two-fluid model in quantum hydrodynamics.}}
%\end{center}
\label{fig:Figure-SchemView}
\end{figure}

Our two-fluid model differs from the ordinary two-fluid model in its derivation and interpretation of the equation. Here, the ordinary two-fluid model is derived from the following microscopic relations. More specifically, the Navier-Stokes equation for an \HL{inviscid} fluid with a temperature gradient term was derived by solving the Gross--Pitaevskii equation, which is a nonlinear $\rm Schr\ddot{o}dinger$ equation, and the Gibbs--Duhem thermodynamic relation in conjunction. This equation describes the dynamics of the superfluid component. In short, the usual two-fluid model is obtained by making ``macro-corrections'' to the microscopic relational equations. The governing equation for the normal flow component was obtained by subtracting the equation for the superfluid component from the \HL{momentum balance equation of the total fluid~\cite{Darve2011}}.
\HL{Figure~\ref{fig:Figure-SchemView}(II) schematically shows the relationship among the Gross--Pitaevskii equation, Gibbs--Duhem thermodynamic relation, and the ordinary two-fluid model.}
In a domain where quantum mechanics is dominant, the superfluid and normal flow components exist independently, and the two components together are considered to conserve mass and momentum (although entropy is said to be carried only by the normal flow component). This microscopic picture was proposed by Tiza and Landau~\cite{TISZA1938, PhysRev.60.356} and was subsequently verified by countercurrent flow experiments. Originally, the two components were considered completely independent; however, experiments have shown that when the heat input is large, the two components exert a mutual frictional force proportional to the velocity difference between them ~\cite{GORTER1949285}.

By contrast, the concept and derivation of our two-fluid model are quite the opposite. 
Notably, we obtain a system of two equations of the same form as in the two-fluid model described above by making particle corrections to the fluid equations, which are macroscopic expressions of relations in the domain dominated by continuum mechanics. 
Figure~\ref{fig:Figure-SchemView}\HL{(I)} presents a schematic of our methodology. 
The details are as follows. 
First, bulk liquid helium at cryogenic temperatures is considered an inviscid fluid. Subsequently, we assume that the viscous fluid component remains at a finite temperature, \HL{as illustrated in Fig.~\ref{fig:Figure-SchemView}(a)}. The amount of the viscous fluid component decreased as the temperature decreased; at absolute zero, it became a completely inviscid fluid. In other words, our two-fluid model assumes the mixing of the fluid components of inviscid and viscous fluids. Thus, the ``fluid particles'' in our model are virtual particles in classical fluid mechanics. Liquid helium, on the other hand, is usually considered an incompressible fluid. However, helium atoms are known to cause Bose-Einstein condensations (BECs) when cooled to the critical temperature (interestingly, although liquid helium is commonly described as a multiparticle system of Bose particles, only 13\% of it actually causes BEC even at absolute zero~\cite{PhysRevLett.49.279}). In any case, the condensate components were no longer fluid. These state changes to and from the condensates are described using quantum statistical mechanics. In other words, from a microscopic viewpoint, the composition ratio of the condensate component to the normal fluid component fluctuates around the average value. Therefore, it can be assumed that the fluid volume fluctuated around the mean value of the equilibrium state. From a macroscopic perspective, the bulk of liquid helium-4 was compressed by a few percent to approximately 13\% in the low-temperature regime, and the volume changed around the average. Accordingly, this physical picture is compatible with the fluid computational model of the explicit SPH method. The explicit SPH method uses a finite-particle approximation of the incompressible Navier-Stokes equations. However, instead of implicitly solving the Poisson equation for pressure, as is typically done for incompressible fluids, this method solves the equation of state of a fluid, for example, the Tait equation of state~\cite{Monaghan_2005}. This approach essentially reproduces density fluctuations~\cite{becker2007weakly, MONAGHAN1994399, NOMERITAE2016156}. An important point is that we assumed weak compressibility for liquid helium. The ability to solve the Navier-Stokes equations for an incompressible fluid while reproducing its weakly compressible properties is a feature of explicit SPH. In summary, in our two-fluid model, liquid helium is considered an inviscid fluid, and it is assumed that the viscous fluid component remains a function of temperature. In other words, it is a mixture of inviscid and viscous fluids. In addition, weak compressibility was observed. Weak compressibility was reproduced using the explicit SPH method.

In our two-fluid model, we introduce the properties of multiparticle systems as a microscopic correction as follows. First, we assume that the ratio of the number of virtual fluid particles in the inviscid and viscous components is equal to that of the quantum mechanical superfluid and normal fluid components calculated using the BEC theory at this temperature. In other words, the ratio of the number of virtual fluid particles in the inviscid and viscous components can be described as $N_{s}/N$ and $N_{n}/N$, where $N_{s}$ and $N_{n}$ are the numbers of helium atoms in the superfluid and normal flow components, respectively, and $N$ is the total number of atoms that satisfy $N_{s} + N_{n} = N$. The respective density ratios were assumed to follow BEC theory. In other words, the densities of the inviscid and viscous fluids normalized by the total density $\rho$ are $\rho_{s}/\rho$ and $\rho_{n}/\rho$, respectively, and they satisfy the relations $\rho_{s}/\rho = N_{s}/N$ and $\rho_{n}/\rho = N_{n}/N$. Thus, $\rho_{s} + \rho_{n} = \rho$. In our model, the virtual fluid particles have the same volume $V_{p}$ regardless of their particle type. Therefore, the density and mass of the fluid particles are proportional and can be expressed as $\rho_{s} = m_{s}/V_{p}$ or $\rho_{n} = m_{n}/V_{p}$, where $m_{s}$ and $m_{n}$ are the masses of the fluid particles in the inviscid and viscous fluid components, respectively. Under this physical interpretation, the total density of liquid helium remains at $\rho$ because the total mass $M$ of the fluid remains constant at any temperature and the total volume $V$ is also constant. Thus, it is reasonable to solve the system of equations for an incompressible fluid (although the use of explicit SPH also reproduces weak compressibility, as described above). The key points of our model are as follows. Let the mass of the total fluid be $M$, the density $\rho$, and the volume $V$; $\rho = M/V$. Let $\rho_{s}$ be the density of the inviscid component and $\rho_{n}$ be the density of the viscous component. Let $M_{s}$ be the total mass of the inviscid component, and $M_{n}$ be the total mass of the viscous component of the entire fluid. The relationship $\rho_{s} = M_{s}/V$ or $\rho_{n} = M_{n}/V$ holds for the entire fluid. From the relation $\rho = \rho_{s} + \rho_{n}$~$=$~$(M_{s} + M_{n})/V$~$=$~$M/V$, $M_{s} + M_{n} = M$. Accordingly, the mass $M$ is conserved throughout the system, and its component ratio is determined by the temperature $T$. The volume of the fluid is also always $V$ since the total density $\rho$ is always constant. Even if we correspond the composition ratio of the fluid particles to that of the actual microparticles, the continuum nature of the fluid is not lost.

Although the two components may exhibit particle properties as described above, in principle, they are ``fluid components,'' and therefore, an interface between the two components is considered to exist under certain conditions. Even if the temperature of the fluid is constant, the density difference causes a temperature gradient between the two fluid particles near the interface, which acts as interfacial tension. In summary, an interfacial tension proportional to the temperature gradient is generated in the opposite direction between the two fluid components, as shown in \HL{Fig.~\ref{fig:Figure-SchemView}(b)}. In summary, the inviscid component follows the Euler equation with a temperature gradient term, whereas the viscous component follows the incompressible Navier--Stokes equation with a temperature gradient term in the opposite direction \HL{(Figs.~\ref{fig:Figure-SchemView}(a) and (b))}. At this stage, only the temperature gradient force contributes to the interfacial tension between the two components. Temperature is a macroscopic state quantity. In other words, only the macroscopic physical properties contribute to interfacial tension. However, as the temperature approaches absolute zero, more local interactions between the two components are expected to contribute to interfacial tension. In other words, local properties have emerged as multiparticle systems. Therefore, in addition to the macroscopic temperature gradient force, the local interaction force $\vec{F}_{sn}$ between two fluid particles is explicitly described as part of the interfacial tension. In classical fluid mechanics, each phase is subjected to an action proportional to the local velocity difference between the different phases at the interface of a two-phase flow. In the Lagrangian picture, the action force between two different fluid particles is proportional to the local velocity difference between the two components. More specifically, as shown in \HL{Fig.~\ref{fig:Figure-SchemView}(c)}, the $i$th fluid particle is subjected to a force proportional to the velocity difference between its own velocity and the average of the velocities of different types of fluid particles, and its reaction force is reflected in the different types of fluid particles. Because we use the explicit SPH method described above, the interface model of the two-layer flow of the explicit SPH method can be used as is. 
We examined the conditions under which the interface between the two phases should be considered. First, the ``interface'' between the two phases is a fundamental concept in hydrodynamics. Therefore, the interface effect becomes non-negligible when the two components exhibit hydrodynamic rather than many-body properties. Liquid helium-4 is expected to behave as a uniform inviscid \HL{fluid} when no external force is applied. However, the assumption of a single-phase flow of an inviscid fluid is broken by a strong external shock or heat, and an interface effect between the two components appears. As this situation simultaneously reveals the multi-particle nature of the fluid, the interfacial effect operates only in the direction of the applied impact force, and the force does not propagate in other directions as in a continuum. In the horizontal rotation problem, liquid helium-4 was forced to rotate horizontally; therefore, fluidity with weak compressibility may manifest strongly in the horizontal direction. Accordingly, the driving force due to the temperature gradient force was assumed to act in the horizontal direction. Interface effects in \HL{the vertical} direction were not considered in our model because no external forces acted in \HL{this} direction.

The above can be summarized as follows. Our two-fluid model assumes mixing of inviscid and viscous fluid components. The inviscid component follows the Euler equation with a temperature gradient term, whereas the viscous component follows the incompressible Navier-Stokes equation with a temperature gradient term in the opposite direction. The interface effect between the two components owing to a temperature gradient acts in the induced direction only when classical fluidity is induced between the two components by excitation or external forces. We consider the local interaction force $\vec{F}_{sn}$ acting between two different types of fluid particles that is proportional to the velocity difference between the different types of fluid particles. In addition, \HL{as shown in Fig.~\ref{fig:Figure-SchemView}(d)}, the viscous term was reformulated to conserve the angular momentum of the fluid particles around their axes, that is, the spin angular momentum. \HL{This is explained further later.} \HL{We admit weak compressibility under specific conditions, and therefore, adopt an explicit SPH as in Fig.~\ref{fig:Figure-SchemView}(e)}. The resulting system of equations is as follows~\cite{Tsuzuki_2021, doi:10.1063/5.0060605, doi:10.1063/5.0122247}
\begin{eqnarray}
\rho_{s} \frac{{\rm D} \vec{v}_{s}}{{\rm D} t} &=& -\frac{\rho_{s}}{\rho}\nabla P + \rho_{s}\sigma\nabla T - \vec{F}_{sn}, \label{eq:goveqsuper:mut}\\
\rho_{n} \frac{{\rm D} \vec{v}_{n}}{{\rm D} t} &=& -\frac{\rho_{n}}{\rho}\nabla P - \rho_{s}\sigma\nabla T + (\bar{\eta} + \bar{\eta_{r}})\nabla^2 \vec{v}_{n}  
		+ \biggl( \frac{\bar{\eta}}{3} + \bar{\xi} -\bar{\eta_{r}} \biggr) \nabla\nabla\cdot\vec{v}_{n} + 2\bar{\eta_{r}}\nabla\times\vec{\omega}_{0} + \vec{F}_{sn}. \label{eq:goveqnormal:mut}
\end{eqnarray}
As shown in Eqs.~(\ref{eq:goveqsuper:mut}) and (\ref{eq:goveqnormal:mut}), the mathematical expressions for the \HL{five} assumptions shown in Figs.~\ref{fig:Figure-SchemView}(a)--(e) can be described in a form similar to that of the conventional two-fluid model derived from the $\rm Schr\ddot{o}dinger$ equation. \HL{Let us explain this further. Figure~\ref{fig:Figure-SchemView}(I) illustrates the assumptions (a)--(e) that derive our two-fluid model. Equations (h) and (j) in Fig.~\ref{fig:Figure-SchemView} correspond to Eqs.~(\ref{eq:goveqsuper:mut}) and (\ref{eq:goveqnormal:mut}), respectively. The underlined subscripts in each term of the equations indicate the assumptions underlying the introduction of these terms. For example, Eq.~(h) is the Euler equation for an inviscid fluid with a temperature gradient and is based on assumptions (a)--(c). Equation (i) can be derived based on assumptions (a)--(c), and it can also be obtained by subtracting Eq. (h) from the momentum balance equation for the total fluid~\cite{Darve2011}. In addition, we reformulate the viscosity term in Eq. (i) to conserve the rotational angular momentum of each constituent particle, as described below. For comparison, the ordinary two-fluid model, which was derived from the Gross--Pitaevskii and Gibbs--Duhem thermodynamic equations, is shown in (k) and (l). 
As previously mentioned, Eq. (l) can be obtained by subtracting Eq. (k) from the momentum balance equation for the total fluid~\cite{Darve2011}. In other words, the only difference between the derivation of the two-fluid model in (I) classical and (II) microscopic depictions is whether the existence of the inviscid fluid is a precondition or whether the inviscid flow is also derived from the microscopic equations and thermodynamic relations. However, the importance of these differences in interpretation has a significant impact on subsequent modeling; this is discussed in section~\ref{sec:challengeofmodel}.	
To summarize, it can be seen from Fig.~\ref{fig:Figure-SchemView} that the governing equations for cryogenic liquid helium are expressed in similar forms both from the classical fluid or quantum mechanical perspectives.}

\HL{In this manner}, we \HL{adopted} the following assumptions. (i) It is a mixture of a viscous and an inviscid fluid \HL{at finite temperature}. (ii) The ratio of helium atoms in the ground state to the total number of atoms follows quantum statistical mechanics for bosons and thus fluctuates around its statistical average, enabling us to assume that the volume of the fluid fluctuates; in other words, weak compressibility can be assumed. (iii) Local interaction forces are exerted between two different types of fluid particles. Moreover, in another recent two-fluid model, the vortex filament model~\HL{\cite{Idowu2001, PhysRevLett.120.155301, doi:10.1063/1.5091567, PhysRevLett.124.155301}} was solved instead of the inviscid equation, in which the rotation of the vortices can be considered. In contrast,  (iv) we assume spin angular momentum conservation of fluid particles around their axes; thus, the viscosity term is decomposed into three terms in Eq.~(\ref{eq:goveqnormal:mut}).
\HL{The aforementioned (i) corresponds to Figs.~\ref{fig:Figure-SchemView}(a) and (b), (ii) corresponds to (e), (iii) corresponds to (c), and (iv) corresponds to (d). It should be emphasized that this paper is the first to clearly describe that cryogenic liquid helium-4 is a system in which viscous fluid components ``remain'' in the inviscid fluid at finite temperatures, and that the residual components decrease with decreasing temperature, becoming a ``uniform inviscid fluid'' at absolute zero.}

Let us summarize the physical significance of each parameter in Eqs.~(\ref{eq:goveqsuper:mut}) and (\ref{eq:goveqnormal:mut}) with reference to the literature~\cite{doi:10.1063/5.0060605}. $\rho_{n}$ and $\rho_{s}$ are the mass densities of the normal and inviscid fluid components, respectively, which satisfy the relationship $\rho = \rho_{n} + \rho_{s}$, where $\rho$ is the mass density of the total fluid. $D\{\cdot\}/Dt$ denotes the material derivative. $\vec{v}_{n}$ and $\vec{v}_{s}$ are the velocities of normal and inviscid fluid components, respectively. $P$, $T$, and $\sigma$ denote the pressure, temperature, and entropy density, respectively. $\bar{\eta}$, $\bar{\xi}$, and $\bar{\eta_{r}}$ indicate the shear, bulk, and rotational viscosities, respectively, where $\bar{\eta_{r}}$ determines the impact of the rotational forces~\cite{MULLER2015301}. 
\HL{The rotational force here indicates the rotational force around the axis of each fluid particle. The vector $\vec{\omega}_0$ in the fifth term on the right-hand side of Eq.~(\ref{eq:goveqnormal:mut}) represents the angular velocity of each fluid particle along its axis. While the vorticity is defined on a continuum, the vector $\vec{\omega}_{0}$ is a variable of the microscopic constituent particles, which in its original formulation represents molecular particles and later approximates fluid particles. Hence, the vorticity and $\vec{\omega}_0$ are different variables. This is explained further. Originally, Condiff~\cite{doi:10.1063/1.1711295} reformulated the Navier--Stokes equations from Cauchy's equation of motion and the conservation law of angular momentum to include the internal freedom of molecular particles to account for the effect of their spins on the entire polar fluid, resulting in a Navier--Stokes equation with spin-angular momentum conservation. Because molecules are polarized in polar fluids, the influence of the rotational motion of individual molecules on the bulk fluid needs to be considered. A schematic of the derivation of this equation is presented in the Appendix Fig.~\ref{fig:Figure-NSeqWithSpinConsv}. It is worth noting that Condiff divided the total angular momentum ($\vec{M}$) per unit mass of the fluid into the angular momentum of the bulk fluid ($\vec{r} \times \vec{u}$) and the contribution due to the internal degrees of freedom of the molecules, that is, the rotational angular momentum ($\vec{I}$) (see Fig.~\ref{fig:Figure-NSeqWithSpinConsv}(D)). The vectors $\vec{r}$ and $\vec{u}$ represent the coordinates and velocity of the local fluid fragment, respectively. $\vec{I}$ can be further decomposed into $\vec{\it I}\cdot \vec{\omega}_{0}$, where $\vec{\it I }$ is a tensor field, which is a scalar multiple of the unit dyadic if we postulate a uniform and isotropic rotational field (and thus a rigid-body rotation) for each constituent particle.
In this context, $\vec{\omega}_{0}$ denotes the angular velocity of a molecule's rotation around its axis. Thus, $\vec{\omega}_{0}$ is in principle a quantity defined for a molecule. However, $\vec{\omega}_{0}$ is often redefined for each fluid particle comprising a collection of molecules, assuming that each fluid particle is a representative point among the subordinate molecules based on a coarse-grained approximation. Therefore, the term ``fluid particle'' here is closer to the definition in the context of molecular fluid dynamics, and thus, may be not the same as a fluid particle as defined in classical fluid mechanics. When considering mesoscale flow problems, such as molecular fluids, fluid particles are often interpreted in both the classical and microscopic senses; we still refer to them as classical to compare the quantum mechanical description of the ordinary two-fluid model in this paper.

In the Lagrangian picture, the difference between the vorticity $\nabla \times \vec{u}$ owing to the rotational motion of the bulk fluid and the vorticity owing to the rotation of the molecules can be assumed to cause strain. Specifically, we assume that the deviation of the vorticity from the rigid-body rotation yields strain because the molecules rotate rigidly. When the vorticity $\nabla \times \vec{u}$ at a point is a rigid-body rotation, the local rotation of the fluid is synchronized with the spin of the molecule; thus, $\nabla \times \vec{u}-2\vec{\omega}_0 = 0$. In this case, no viscous stress owing to local rotation occurs. Otherwise, the difference in $\nabla \times \vec{u}-2\vec{\omega}_0$ induces strain in the fluid, making it that deviate from rigid-body rotation, and contributes to the asymmetric part of the stress tensor (see Figs.~\ref{fig:Figure-NSeqWithSpinConsv}(F) and (I) in the Appendix). 
In the third through fifth terms on the right-hand side of Eq.~(\ref{eq:goveqnormal:mut}), the viscosity term is divided into three contributions: shear, volume, and rotational viscosities. Thus, the spin-angular momentum-conserving Navier--Stokes equations can be rephrased as Navier--Stokes equations with reformulated viscosity terms. This model can easily connect microscopic and fluid mechanics in a continuum mechanical regime. To date, it has been applied to mesoscale flows in which the rotation of molecules or fluid particles dominates, such as in polar fluids and suspension flows. Previously, M{\"u}ller adopted SPH to discretize the spin-angular momentum-conserving Navier--Stokes equations to simulate two immiscible flow~\cite{MULLER2015301}; their model was further applied to red blood cell flow~\cite{doi:10.1073/pnas.1608074113}. 

Mathematically, SPH is a finite particle approximation of a continuum field that can decompose the quantities defined in the field into neighboring particle contributions. The intrinsic part of SPH is a representation of a physical quantity in an integral form using the Dirac delta function $\delta$ and its approximation by the finite distribution function $W$, called the smoothed kernel function. We then discretize the approximation form based on the concept of the sum approximation for numerical simulations. This paper does not focus on the derivation of the SPH models because they have been discussed in our previous studies~\cite{Tsuzuki_2021, doi:10.1063/5.0060605, doi:10.1063/5.0122247}; however, the basic concepts of SPH and its discretized models used in this study are illustrated in Fig.~\ref{fig:Figure-SchemOfSPH} in the Appendix. In Refs.~\cite{Tsuzuki_2021, doi:10.1063/5.0060605, doi:10.1063/5.0122247}, the viscosity term of the two-fluid model for cryogenic liquid helium-4 was reformulated to conserve the spin-angular momentum. In other words, the viscosity term was reformulated so that the two-fluid model can be applied not only to shear viscosity-dominated problems such as countercurrent flow, but also to rotation-dominated problems. We then discretized the equations using the SPH. Consequently, we observed the formation of multiple spinning vortices that occurred independently of the overall fluid motion in our SPH simulations of the rotating 2D liquid helium-4~\cite{Tsuzuki_2021}. After incorporating the vortex dynamics model into our model, we found that multiple spinning vortices formed a lattice with a regular arrangement~\cite{doi:10.1063/5.0060605, doi:10.1063/5.0122247}. The number of vortices generated was determined based on the strength of the rotational angular velocity. For appropriate values, the number of vortices agreed well with the theoretical solution of the number of vortices calculated using Feynman's law~\cite{doi:10.1063/5.0060605}. Compared with our previous studies, this study is the first to address a 3D problem. The system does not assume gravity in any direction, and considering numerical stability, the validation in the 2D cases is effective for the 3D cases as well. As we focus on the disk-shaped problem and consider only the horizontal rotation around the vertical axis (z) of the disk, the developed rotational viscosity term for the two-dimensional model can be used directly. However, because of the topological differences in vortices between the 2D and 3D cases, it needs to be studied whether the previous method, which is effective in solving the 2D rotating problem, can numerically reproduce the vortices in 3D liquid helium-4; large-scale simulations for 3D cases will be reported and discussed later.}

The fourth term on the right-hand side of Eq.~(\ref{eq:goveqnormal:mut}) becomes $\rm 0$ owing to the incompressibility condition $\nabla \cdot \vec{v} = 0$. In this respect, the current model always imposes the incompressibility condition $\nabla \cdot \vec{v} = 0$, and the simulation is slightly weakly compressible by solving using the explicit SPH method. Therefore, at present, the fourth term is meaningless; however, in the future, it may be solved directly under a compressibility condition. Therefore, we describe the system of equations in a generic form as in Eq.~(\ref{eq:goveqnormal:mut}).
The viscosity term in Eq.~(\ref{eq:goveqnormal:mut}) is an extended version of the usual expression: the set of the third to fifth terms converges to $\bar{\eta}\nabla^2 \vec{v}_{n}$ because $\bar{\eta_{r}}$ converges to $0$ when $\nabla \cdot \vec{v} = 0$. This implies that $\bar{\eta}$ corresponds to the shear viscosity $\eta_{n}$ of the normal fluid. 
Meanwhile, the local interaction force $\vec{F}_{sn}$ is expressed as follows. As mentioned earlier, $\vec{F}_{sn}$ is proportional to the local velocity difference between the two components, and can be expressed as $\vec{F}_{sn} = C_{l} \vec{v}_{sn}$, where $C_{l}$ is a coefficient. This form exactly corresponds to the mutual friction forces between the two components in quantum mechanics; in this paper, we estimate $C_{l}$ assuming that $\vec{F}_{sn}$ corresponds to the mutual friction forces $\vec{F}^{q}_{sn}$ of the ordinary two-fluid model, where $\vec{F}^{q}_{sn}$ can be expressed as $\vec{F}^{q}_{sn} = 2/3 \rho_{s} \alpha \kappa L \vec{v}_{sn}$, where $\vec{v}_{sn}$ is the relative velocity $\vec{v}_{s} - \vec{v}_{n}$, $\kappa$ is the quantum of circulation, $\alpha$ is the friction coefficient, and $L$ is the time-dependent vortex line density. Unlike in our previous studies, we use the initial value of $L$ before its decay, $L^{-1}(0)$, where $L$ satisfies Vinen's equation: $L^{-1}(t) = L^{-1}(0) + \beta_{v} t,$ where $\beta_{v}$ is a coefficient. For further details, refer to~\cite{NEMIROVSKII201385}. 

In addition, two auxiliary equations are required to solve the system of equations in Eqs.~(\ref{eq:goveqsuper:mut}) and (\ref{eq:goveqnormal:mut}) such that the number of unknown independent variables equals the number of equations. Specifically, the elementary excitation model~\cite{Adamenko_2008, schmitt2015introduction} establishes a correlation between the entropy $\sigma$ and temperature $T$. Tait's equation of state describes the barotropic relationship between the pressure $P$ and density $\rho$ for each fluid particle~\cite{Monaghan_2005}:
\begin{eqnarray}
\sigma &\simeq& \frac{1}{NM}\Biggl[\frac{2\pi^2 k_{B}^4 T^{3}}{45 \hbar^3 c^{3}} + \biggl(\frac{k_{B}}{2\pi}\biggr)^{3/2}\frac{\sqrt{\mu}p_{0}^2 \Delta}{\hbar^3}\frac{e^{-\Delta/T}}{\sqrt{T}}\Biggr], \label{eq:goldentro} \\
P &=& p_{0}\Biggl(\frac{\rho}{\rho_{0}}\Biggr)^{\alpha_{P}} - \beta_{P}. \label{eq:prtcpres}
\end{eqnarray}

Let us explain the physical significance of each parameter in Eqs.~(\ref{eq:goldentro}) and (\ref{eq:prtcpres}) with reference to Refs.~\cite{Tsuzuki_2021, doi:10.1063/5.0060605}.
In Eq.~(\ref{eq:goldentro}), $M$ represents the mass of a helium-4 atom, $N$ indicates the total number of helium-4 atoms in the system, $\pi$ represents \HL{the ratio of the circumference of a circle to its diameter}, and $c$ is the speed of sound. $\hbar$ indicates the reduced Planck's constant and $k_{B}$ represents the Boltzmann constant. The constant values $\mu$, $p_{0}$, and $\Delta$ are given by ${\rm 1.72\times10^{-24}~g}$, ${\rm 2.1\times10^{-19}~gcms^{-1}}$, and ${\rm 8.9~K}$, respectively ~\cite{schmitt2015introduction, bennemann2013novel}.
Eq.~(\ref{eq:goldentro}) is always true within the scope of this study because it is valid if $T \ll {\rm 93~K}$~\cite{schmitt2015introduction}.
In the elementary excitation model, the first term in the parentheses on the right side of Eq.~(\ref{eq:goldentro}) represents the ``photon'' contribution, and the second term in the same parentheses represents the ``roton'' contribution; after finding the total pressure from the summation of the photon and roton contributions, the entropy density in Eq.~(\ref{eq:goldentro}) is obtained by the thermodynamic relation, such that the partial derivative of pressure with respect to temperature~\cite{Tsuzuki_2021}.
Equation~(\ref{eq:goldentro}) represents the relationship between entropy density $\sigma$ and temperature $T$ in a quantum ``ideal gas.'' To obtain the value of liquid helium-4, we decompose $\sigma$ as $\sigma=C_{e}\sigma_{0}$, where $C_{e}$ indicates the volume ratio of liquid to gaseous helium-4 under an atmospheric pressure of $\rm 1.01325~bar$, which is $1.428\times 10^{-3}$~\cite{hammond2000elements}. 
Under constant temperature, we calculated $\sigma_{0}$ using Eq.~(\ref{eq:goldentro}) and maintained a constant value for entropy conservation. 
Meanwhile, in Eq.~(\ref{eq:prtcpres}), $\rho_0$ and $p_{0}$ represent the fluid density and pressure in the initial state, respectively. $\alpha_{P}$ is the specific heat ratio that affects system incompressibility. $\beta_{P}$ denotes the reference pressure, which typically corresponds to $p_{0}$. Equations~(\ref{eq:goveqsuper:mut})--(\ref{eq:prtcpres}) represent the governing equations of the system.

\subsection{\HL{The potential and immediate challenges of our two-fluid model}} \label{sec:challengeofmodel}
\HL{Notably, for the dynamics of cryogenic liquid helium-4, both the derivation of the governing equations from the mechanical interpretation in Figs.~\ref{fig:Figure-SchemView}(a)--(e) and quantum mechanics resulted in similar mathematical forms. The first model assumes a two-phase flow model in which the two components are separated and the fluid particles have a volume exclusion effect on each other. Particle correction has been introduced as an exception. Conversely, the second or ordinary two-fluid model assumes that both components coexist in space and that neither type of particle exerts volume exclusion effects on each other, and introduces mutual friction as an exception. In summary, the governing equation can be identical regardless of whether the system comprises separate or coexisting (overlapping) particles. However, separate and overlapping constituent particles lead to different physical meanings, and the former and latter models provide the classical and quantum-mechanical descriptions of the same problem. In particular, the physical meaning of ``particles'' is different; the former indicates classical fluid particles, and the latter represents quantum minute particles. This inspired us to reconceptualize the system in terms of the degree of volume exclusion between the two components, hereafter, referred to as the mutual volume exclusion (MVE) effect. Here, the MVE effect can be rephrased as the effect of classical fluid forces, such as surface forces, body forces, or interface tensions. This eliminates the need to distinguish the physical meanings of the particles between the two models, considering that the MVE is the primary factor causing the particles to separate or overlap.   

Ultimately, the two models can be unified into one model in terms of the MVE effect as a variable. This model has the following advantages. New insights into the connection between classical and quantum mechanics can be obtained by using the MVE effect as a unified metric of the system. This may also provide a new perspective on the relationship between the classical two-phase flow and the ordinary two-fluid model of unified hydrodynamics. In engineering and industry, it may become possible to capture classical and quantum fluid phenomena with only one model by adjusting the degree of MVE of the constituent particles depending on the location and condition. Practically, there is no need to switch computational codes depending on the conditions, and a single computational code can be used to analyze the entire large-scale system.

In such a unified model, the transition from classical to quantum particles can, in principle, be automatically performed using high-resolution computational calculations. This is because as the computational resolution increases, the size of the fluid particles in the classical domain gradually decreases, and the excluded volume effect of individual fluid particles also decreases. In other words, a low computational resolution in a local region implies that the size of the analytic particles at that location is large, and a large analytic particle size implies that the excluded volume effect of the individual particles is large, which is suitable for describing the location where the classical fluid is dominant. Conversely, a high computational resolution in a local region implies that the size of the analytical particles at that location is small, which implies that the excluded volume effect of individual particles is small, which is appropriate for a local region where quantum hydrodynamics dominate. In summary, in the unified model, the convergence of the computational resolution is expected to correspond to the convergence of the physical phenomena. In this case, the introduction of multiresolution particles makes it possible to reproduce the behavior of the fluid in the entire system in a single model.

To this end, the immediate challenge of our approach is to provide evidence that increasing the resolution brings the simulation results closer to the phenomena observed in the quantum-mechanical regime, even when using our classical mechanical two-fluid model with spin-angular momentum conservation.}

\subsection{Multi-GPU computing for large-scale particle simulations}
In high-performance computing, an effective parallel computing algorithm for the simulation of many-particle interacting systems should be chosen based on the characteristics of the target problem, computational scheme, and properties of the parallel computing environment. In general, these problems include gravitational many-body problems~\cite{10.5555/2388996.2389003, Sugimoto1990}, molecular dynamics (MD) problems~\cite{https://doi.org/10.1002/jcc.20289, valiev2010nwchem, Heinecke2015}, Lagrangian fluid analysis using SPH~\cite{10.1007/978-3-031-39698-4_38, OGER20161}, deformation analysis of structures using the mesh-free method~\cite{CHEN1996195, Nayroles1992, https://doi.org/10.1002/nme.1620370205, Lin2020, MikioSakai20202020017}, and powder problems using the distinct element method (DEM)~\cite{doi:10.1680/geot.1979.29.1.47, PARK2021104008, Yan2018, GAN2020258}. The interaction domains and connectivity between particles depend on the target problem. From a computational perspective, gravitational many-body problems and MD problems result in N-body calculations owing to the large interaction range. Consequently, the computational cost of the interaction calculations is $O(N^2)$. Conversely, DEM calculations for powders require interaction calculations of the repulsion and friction between the contacting particles. Therefore, the memory reference cost is more significant than the computational cost when appropriate techniques such as neighbor particle listing and particle data sorting~\cite{GREST1989269, 10.1016/j.cageo.2012.02.028, Tsuzuki:2016:EDL:3019094.3019095} are implemented to reduce the interaction calculation cost to $O(N)$. In contrast, fluid calculations using particle methods, such as SPH, stand at a middle ground between the N-body and DEM calculations in terms of both memory reference and computational cost, even if we use a technique similar to DEM. In a 3D calculation, each particle interacts with approximately a few hundred particles within the interaction domain, which is called the kernel radius. Moreover, the numerical accuracy of the spatial discretization in SPH is typically equivalent to that of the central difference method in the finite difference method (FDM). In addition, the time discretization accuracy was between the first- and second-order accuracies with respect to the resolution, depending on the time-integration schemes. Accordingly, it is imperative to capture the underlying real physics with the highest possible accuracy using as many analytical particles as possible. This is one of the reasons why the field of large-scale simulation of particle methods frequently appears in the ACM Gordon Bell Prize~\cite{doi:10.1177/1094342017738610}.

 In recent years, graphics processing units (GPUs) have been installed in world-class supercomputers as accelerators to improve the computational performance, memory utilization, power efficiency, and other metrics. Multi-GPU computing can speed up simulations and ensure that sufficient memory is available. The development of applications that operate efficiently on GPU supercomputers requires CUDA programming~\cite{4541126}. This also requires making good use of hierarchical memory structures, such as the limited amount of device memory on the GPU board and high-speed shared memory on the chip. The GPU supercomputer consists of many nodes with distributed memory connected by high-speed networks. GPUs are installed on each compute node as computational accelerators. For particle simulations on a memory-distributed system such as a supercomputer, it is efficient to divide the entire computational domain in space and assign a node with accelerators to each divided subdomains. Among particle problems, calculations performed in gravitational many-body and MD are based on long-range interactions. Therefore, the computational cost is dominated by floating-point operations rather than memory accesses, which allows for high execution performance, even on GPU supercomputers with low byte/flops. However, in SPH simulations that compute short-range interactions, most computational time is spent on memory accesses, making it impossible for a low byte/flop supercomputer to achieve sufficient performance. 

The computational cost per GPU for particle simulations using SPH depends on the number of particles in the subdomains to which each GPU is assigned. In a parallel computing platform such as a supercomputer, the distribution of particles among nodes changes in both time and space, resulting in a non-uniform distribution of particles per node. At this point, the computational load in regions that are densely populated with particles increases significantly, making large-scale computations impractical. Therefore, it is essential to introduce dynamic load balancing between nodes to achieve sufficient execution performance in simulations of short-range interaction-based particle methods, such as SPH. This involves repartitioning the computational domain according to particle distribution to ensure that the computational load on each node is uniform. However, achieving dynamic load balancing in simulations of short-range interaction-based particle methods on GPU supercomputers is challenging because of hierarchical memory structures. The implementation is further complicated by the need for specialized technologies specific to large-scale parallel computing, such as CUDA programming and the MPI library~\cite{mpi41}, for the development of the computational code. In addition, when the particle distribution undergoes significant changes and moves in a random direction across nodes, it is essential to efficiently manage the particle data on the GPUs to minimize the communication cost between GPUs across nodes via host CPUs and between GPUs and CPUs within each node. It is critical to consider efficient computation methods for individual GPUs considering their different memory structures, such as the device memory on the GPU and the high-speed shared memory on the chip. 
\begin{figure}[t]
\vspace{-21.3cm}
\hspace{29.1cm}
%\begin{center}
\centerline{\includegraphics[width=4.55\textwidth, clip, bb= 0 0 3948 1381]{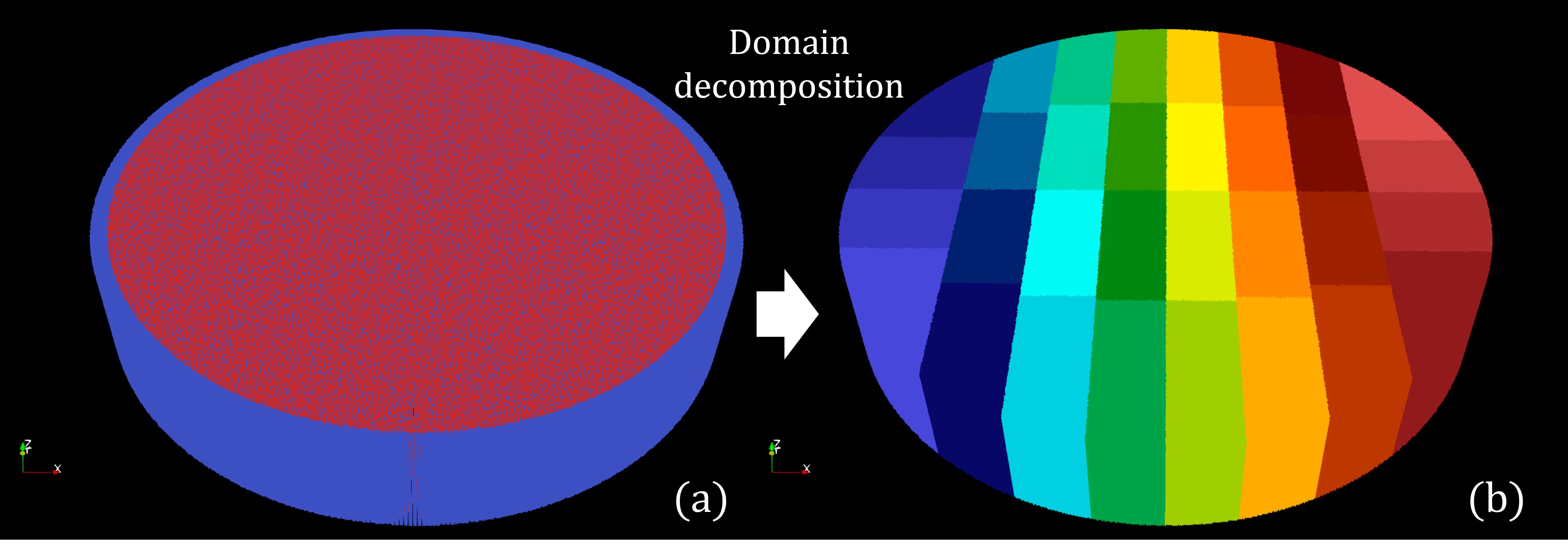}}
%\begin{center}
\caption{\HL{Snapshots of (a) the particle distribution, with inviscid and viscid fluid particles colored in red and blue, respectively, and (b) the domain decomposition in the case where approximately 19.6 million fluid particles were divided into 32 subdomains, each assigned to a single GPU.}}
\label{fig:Figure-DomainDecomp}
\end{figure}

 It is imperative to develop dynamic load balancing among the GPUs to exploit the high computational efficiency of multiple GPUs. Various load balancing techniques have been proposed in related fields, including the slice-grid method, which moves the domain boundary to a spatially partitioned region; graph theory-based partitioning (such as k-means), as represented by ParMETIS~\cite{doi:10.1137/S1064827595287997, Karypis2011} and Zoltan~\cite{ZoltanShortTutorial, 10.1109/5992.988653}; and the hierarchically structured tree method, which uses a recursive structure of cells to partition a domain, as exemplified by kd-tree decomposition~\cite{KDTREE9597006}, orthogonal recursive bisection (ORB)~\cite{Dubinski1996APT, ORBLiu2001, 5333803}, and domain decomposition using a hierarchically structured grid with space-filling curves~\cite{SFCDecomp1997-Aluru}. Recently, we conducted a study on optimal dynamic domain decomposition for particle simulations of short-range interactions compatible with low byte/flops supercomputers, such as GPU supercomputers. Subsequently, we developed a multi-GPU computing framework that enables large-scale particle simulations using SPH. The SPH computational framework~\cite{Tsuzuki:2016:EDL:3019094.3019095} developed by the author was used in this study. Based on the nature of the target problem, the 2D slice grid method was used as the domain decomposition method. Figure~\ref{fig:Figure-DomainDecomp} illustrates a snapshot of the domain decomposition in the case where the 19.6 million fluid particles were divided into 32 subdomains, each of which was assigned to a single GPU to speed up the simulations.

Unlike the free surface problem, the particle distribution in the rotating liquid helium-4 simulation was not excessively biased. Therefore, to prevent the computational load from being excessively biased, we divided the domain at regular intervals, such as every several hundred steps, so that the overhead of domain re-segmentation was negligible, and the load was balanced. A workflow diagram of the calculations is presented in Algorithm~\ref{alg:workflow}. The SPH calculation process for the 3D simulation of liquid helium-4 using our two-fluid model was, in principle, the same as that for the 2D case reported in our previous study. The velocity Verlet method~\cite{PhysRev.159.98, 10.5555/76990} was used as the time integration method, and a nonslip condition was imposed on the lateral wall boundaries. As this is a three-dimensional calculation, a slip condition is imposed on the top and bottom surfaces, assuming that the inviscid fluid continues. After the particle density was calculated, the particle pressure was determined using the Tait's equation of state. After applying the pressure filter~\cite{imoto2019convergence}, the updated particle density and particle pressure were used to calculate the pressure and temperature gradient forces. The viscosity term was separately calculated for the rotational and shear viscosities of the particles. The collision model~\cite{SHAKIBAEINIA201213} computed in line 17 is a correction model that preserves the volume effect of the fluid particles. The local interaction forces between the viscous and inviscid fluid particles were then computed alternately, the boundary conditions were calculated, and the physical quantities of the particles were updated according to the velocity Verlet algorithm. For the detailed description of the computational model using SPH, please refer to Refs.~\cite{Tsuzuki_2021, doi:10.1063/5.0060605}.

\begin{algorithm}[t]
\caption{Computational workflow for simulations using our SPH model on a multi-GPU platform} \label{alg:workflow}
\begin{algorithmic}[1]
{\small
\For{${\tt j}$ = 0 $\to$ ${\tt N_{step}}$}
\State {Re-decompose the domain using the 2D slice-grid method (at a fixed interval).}
\State {Construct Neighbor Particle Lists (NPLs) in each subdomain.}
\For{${\tt k}$ = 0 $\to$ ${\tt N_{velver}}$}
\State {Communicate the halo particles (all particles in the halo).}
\State {Register particles in the NPLs.}
\If{{\tt k}~$=$~{\tt second step}} 
\State {Compute the particle densities.}
\State {Compute the particle pressures.}
\State {Communicate the halo particles (only updated particles in the halo).}
\State {Compute the pressure filter.}
\EndIf
\State {Communicate the halo particles (only updated particles in the halo).}
\State {Compute the pressure gradient forces.}
\State {Compute the temperature gradient forces.}
\State {Compute the viscosity forces.}
\State {Compute collisions. }
\State {Compute local interaction forces (from viscid to inviscid).}
\State {Communicate the halo particles (only updated particles in the halo).}
\State {Compute local interaction forces (from inviscid to viscid).}
\State {Calculate boundary conditions (from fluid to wall particles).}
\State {Communicate the halo particles (only updated particles in the halo).}
\State {Calculate boundary conditions (from wall to fluid particles).}
\State {Calculate time integration (update using Velocity Verlet method).}
\State {Communicate out-of-domain particles.}
\EndFor
\State {Defragment GPU memory (at a fixed interval).}
\EndFor
}
\end{algorithmic}
\end{algorithm}

Given the significant increase in the computational cost associated with 3D calculations, it is imperative to run simulations on a multi-GPU platform. As previously mentioned, dividing the computational domain into subdomains and assigning a GPU to each subdomain is a highly efficient approach. To ensure the connectivity of the domains between subdomains, it is necessary to copy the updated particle data to the neighboring subdomains each time the particles are in the halo region, which is between the boundaries with the neighboring domain and the interior of the domain by the interaction length (in our implementation, twice the kernel radius), update their properties. As shown in Algorithm~\ref{alg:workflow}, five communications of particle data within the halo domain were required. In addition, the particle positions are updated after all computations in a single step were completed. Each GPU then transfers the out-of-domain particles that have left the subdomain to the target neighboring subdomain.  Consequently, communication between neighboring GPUs was required six times. Frequent repetition of communication fragments the particle array and reduces the performance; therefore, periodic reordering of particle data is performed. This process enables efficient multi-GPU computation using SPH.

\begin{figure}[t]
\vspace{-35.0cm}
\hspace{27.0cm}
%\begin{center}
\centerline{\includegraphics[width=4.21\textwidth, clip, bb= 0 0 5644 3524]{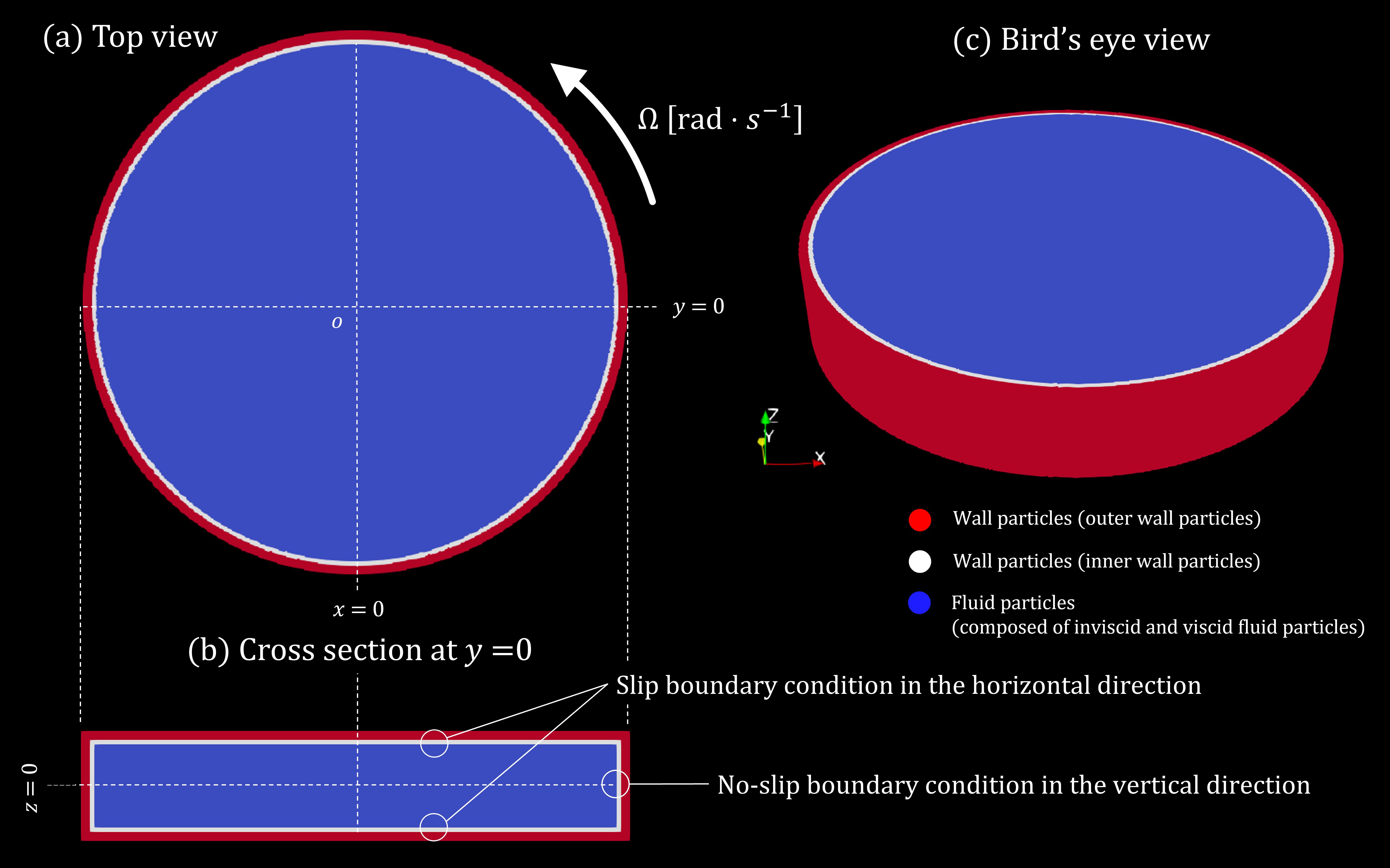}}
\caption{\HL{Geometrical setup of a horizontally rotating 3D liquid helium-4: (a) top view, (b) cross section at y = 0, and (c) bird's eye view.}}
%\end{center}
\label{fig:Figure-GeomSetup}
\end{figure}

\section{Large-scale simulations of horizontally rotating 3D liquid helium-4}
Let us set the outer diameter of a cylindrical container in the horizontal direction to 0.2 cm with a thickness of 0.04 \HL{cm} in the vertical direction (z-direction). The resolutions of $(N_{x}, N_{y}, N_{z})$ are given by (500, 500, 100), where $N_{a}$ $(a = x, y, z)$ indicates the number of particles in the direction in a rectangular domain that covers the cylindrical container. 
We set the rotational angular velocity around the cylinder axis (z axis) to 5 rad $\cdot s^{-1}$ counterclockwise. 
Similar to the two-dimensional cases in our previous work, we set the temperature T to 1.6 K and determined the values of $(\rho_{s}/\rho, \rho_{n}/\rho, \kappa)$ with those at T = 1.6 K listed in Ref.~\cite{Vinen2002}. 
In addition, $\eta_{n}$ was obtained from the relationship of 0.566 $\kappa \rho_{n}$ with reference to~\cite{Vinen2002}. The shear viscosity $\bar{\eta}$ and rotational viscosity $\bar{\eta_{r}}$ have the value of $\eta_{n}$. We set the parameter $C_{\omega}$, which determines the magnitude of the angular velocity $\vec{\omega}_0$ and thereby determines the resulting number of vortices, to be 0.016 to maintain a similar degree of magnitude of the rotational forces as those in 2D cases reported in our previous studies. Normal fluid particles were placed along the inner circumference of the container as walls, and no-slip conditions were imposed on the wall particles as boundary conditions. Conversely, we set slip conditions at the top and bottom of the vessel. \HL{We applied a well-established wall and dummy particle method~\cite{XU2013101, XU201643}; we update the density and pressure of the inner wall particles similarly to fluid particles, and retain the initial pressure and density values of the outer wall particles.} \HL{Figure~\ref{fig:Figure-GeomSetup} illustrates the geometrical setup of the target system.} In the initial state, superfluid or normal fluid particles were randomly generated in proportion to the density ratio and distributed in the vessel. 

\HL{The computational conditions in this study were adopted from the parameter values determined in our previous study~\cite{Tsuzuki_2021}. In addition to satisfying the Courant-Friedrichs-Lewy (CFL) condition, several benchmark problems were solved to validate the stability and numerical accuracy of the calculations. Specifically, we simulated the propagation of the wave equation using the proposed SPH method and compared its numerical solution with the numerical and theoretical solutions obtained using the difference method. In addition, we compared the simulation results with the exact solution for the case in which the Reynolds number $R$ of the normal flow component was 100, and for the case in which $R$ was 1000 in 2D driven flow problems. We obtained very good agreement between the simulation results and the exact solutions obtained by the high-order FDM~\cite{GHIA1982387}. Furthermore, assuming an ideal case in which the two components behave as perfect classical fluids, we solved the Rayleigh-Taylor instability problem under conditions where the density difference $(\rho_{s}/\rho_{n})$ was five times, which is the same as that for the problem covered in this study, and confirmed that the characteristic mushroom structure could be reproduced. These results were confirmed for 2D problems. Although we targeted a 3D problem, we did not need to impose gravity on the system. In addition, the target problem was quasi-2D rather than symmetric 3D, and dynamic changes in the vertical direction were not dominant. Accordingly, we used the values used in previous simulations. Note that we set $\delta t$ to be sufficiently small to satisfy the CFL condition for all resolution cases.}
We set $\delta t$ to $5.0 \times 10^{-6}$ s and computed 180000 time steps, which is 0.9 s in physical time. In the simulation, under the aforementioned conditions, the total number of normal and inviscid fluid particles, including the wall particles, was 19636400. We boosted our simulation using 32 GPUs installed on the Wisteria/BDEC-01 supercomputer at The University of Tokyo, JAPAN. We applied a two-dimensional slice-grid technique to the dynamic domain decomposition in the horizontal direction and decomposed the computational domain into 32 subdomains (four in the horizontal direction and eight in the depth direction), and a single GPU was assigned to each subdomain. Consequently, approximately 5.5 days were required to compute all 180000 time steps of the simulation. For comparison, we performed simulations with a lower resolution using $(N_{x}, N_{y}, N_{z})$ = (250, 250, 50) to investigate the dependence of the observed phenomena on the computational resolution.

\begin{figure}[t]
\vspace{-51.9cm}
\hspace{28.9cm}
%\begin{center}
\centerline{\includegraphics[width=4.5\textwidth, clip, bb= 0 0 3317 2897]{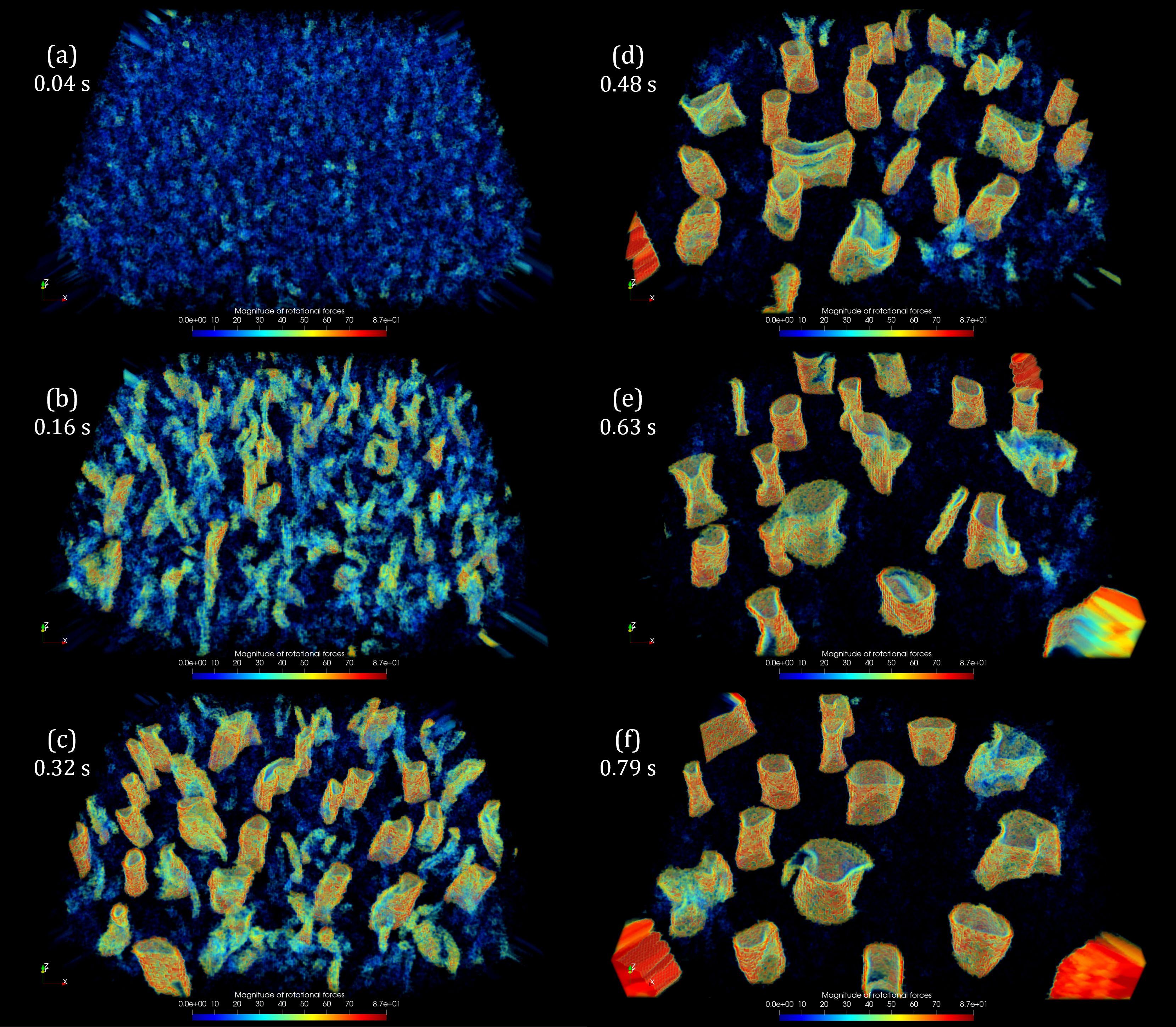}}
\caption{Snapshots of the rotation simulation with the lower resolution using 2454000 particles from the beginning to 0.79 s in physical time.}
%\end{center}
\label{fig:Figure-SimLowReso}
\end{figure}

\begin{figure}[t]
\vspace{-0.5cm}
%\hspace{28.9cm}
%\begin{center}
\centerline{\includegraphics[width=1.0\textwidth, clip, bb= 0 0 2856 1662]{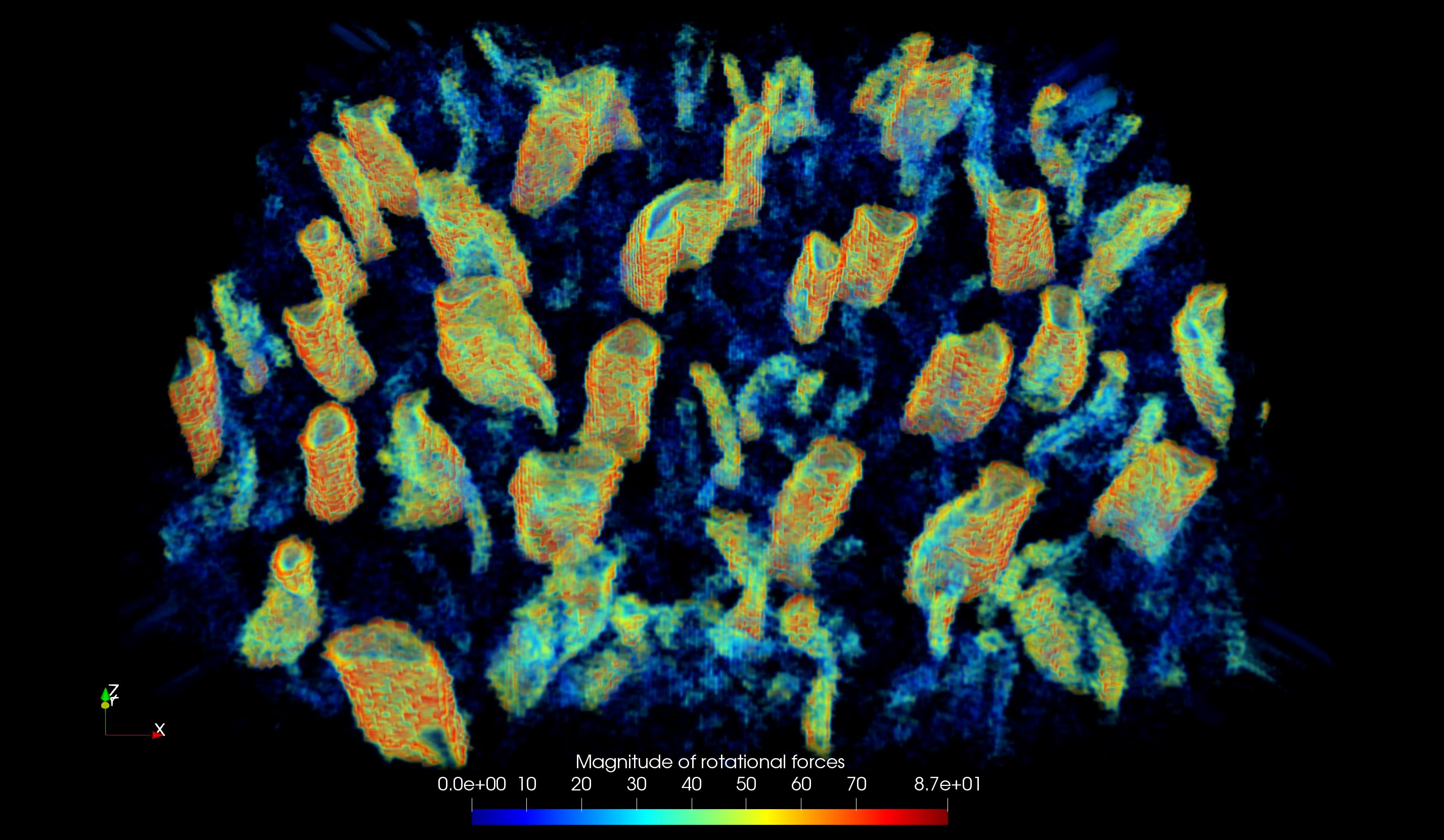}}
\caption{A video of the snapshots in Fig.~\ref{fig:Figure-SimLowReso} (Multimedia view)}
%\end{center}
\label{fig:Figure-Video-LowResSim-0160}
\end{figure}

\begin{figure}[t]
\vspace{-51.9cm}
\hspace{28.9cm}
%\begin{center}
\centerline{\includegraphics[width=4.5\textwidth, clip, bb= 0 0 3317 2900]{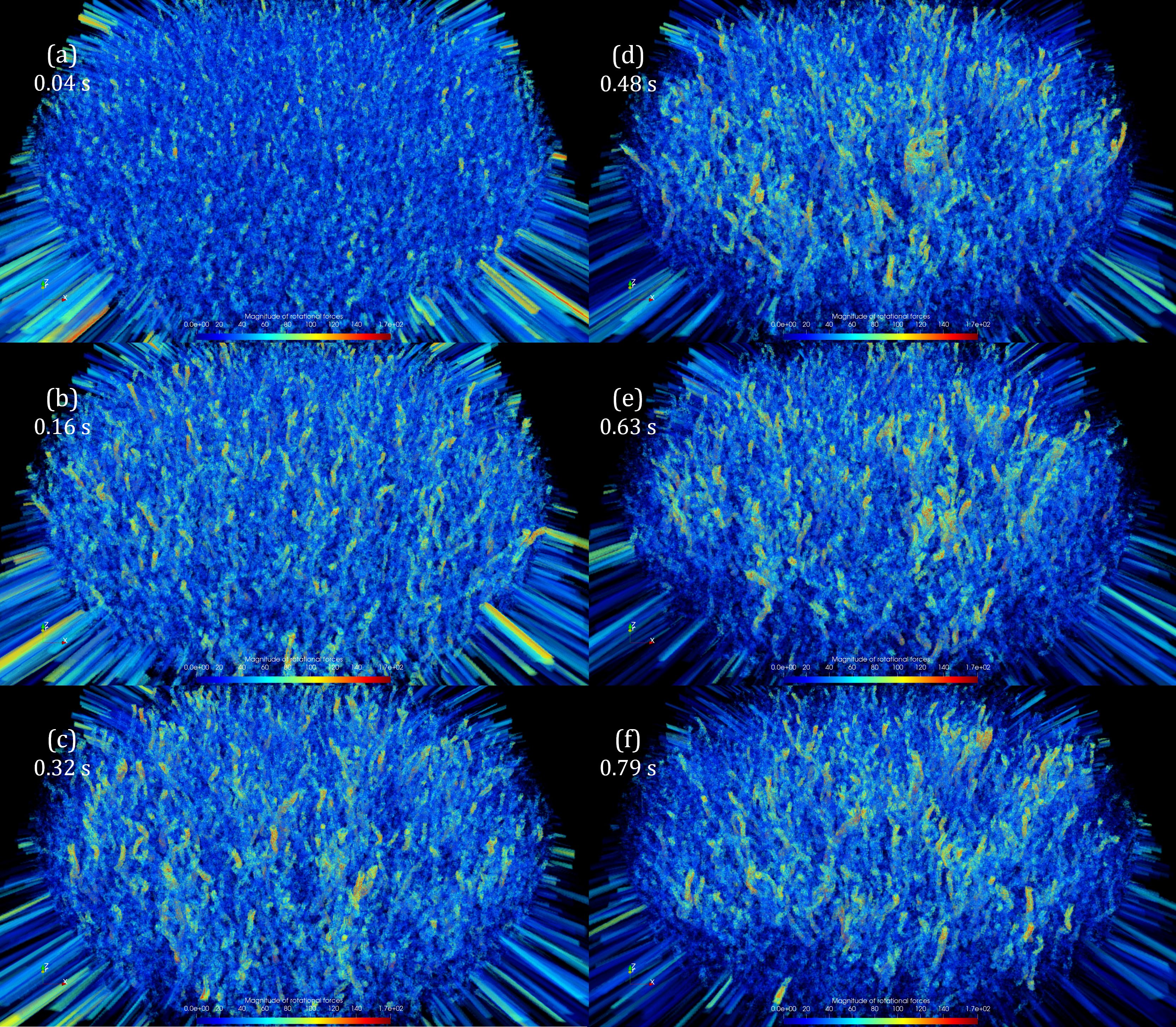}}
\caption{Snapshots of the rotation simulation with the higher resolution using 19636400 particles from the beginning to 0.79 s in physical time.}
%\end{center}
\label{fig:Figure-SimHighReso}
\end{figure}

\begin{figure}[t]
\vspace{-0.5cm}
%\hspace{28.9cm}
%\begin{center}
\centerline{\includegraphics[width=1.0\textwidth, clip, bb= 0 0 2856 1662]{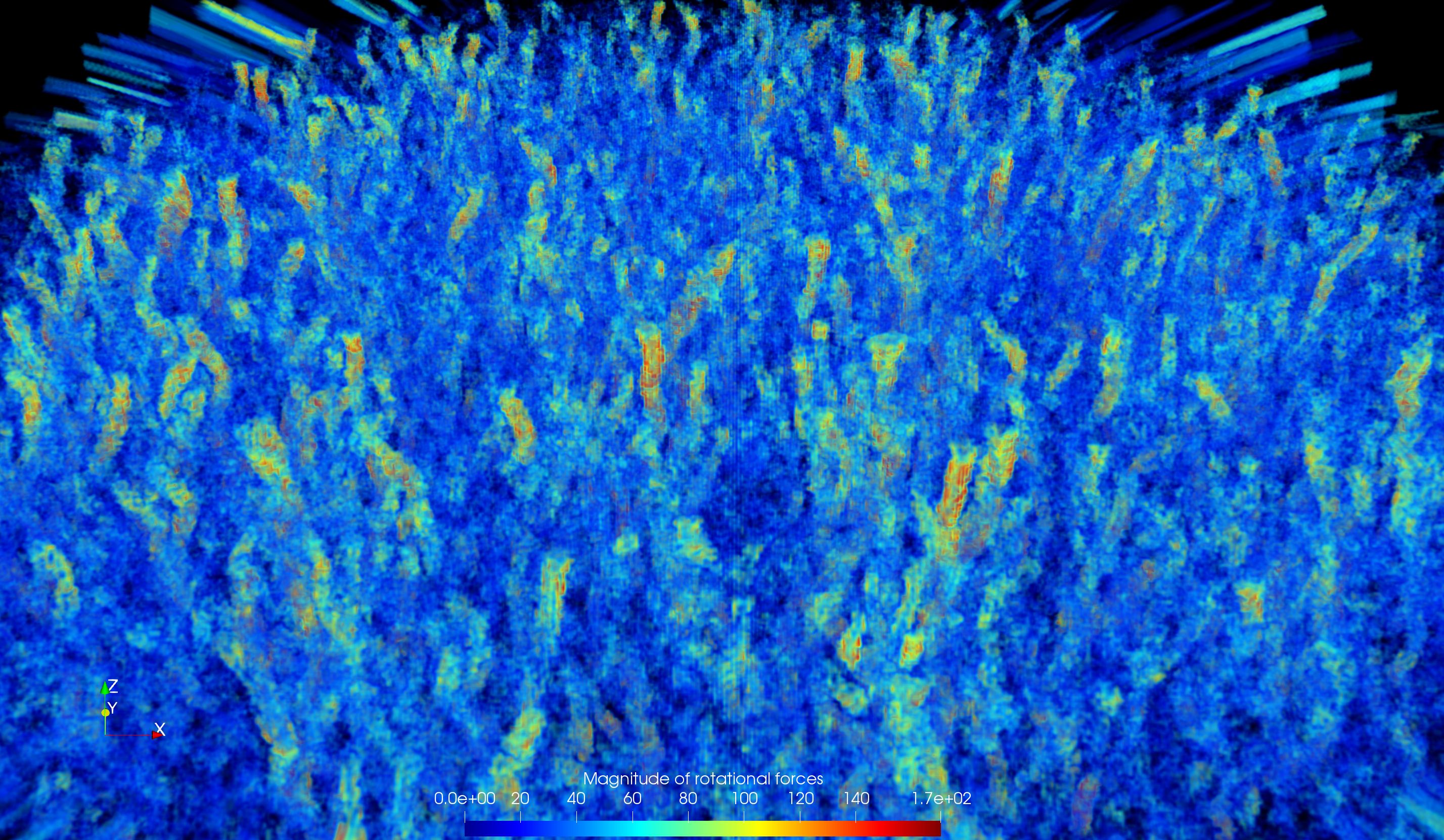}}
\caption{A video of the snapshots in Fig.~\ref{fig:Figure-SimHighReso} (Multimedia view)}
%\end{center}
\label{fig:Figure-Video-HighResSim-0160}
\end{figure}

Figure~\ref{fig:Figure-SimLowReso} shows snapshots of the rotation simulation with a lower resolution using 2454000 particles from the beginning to 0.79 s in physical time. 
The fluid particles of the inviscid and viscous components were visualized in color according to the strength of the rotational forces calculated by the fifth term of Eq.~(\ref{eq:goveqnormal:mut}) using the volume rendering technique.
The wall particles were removed for visualization. A video of the simulation is presented in Fig.~\ref{fig:Figure-SimLowReso} as integral multimedia in Fig.~\ref{fig:Figure-Video-LowResSim-0160}.
A preprint version of the video is available at \\ (\url{https://www.satoritsuzuki.org/video-horizontally-rotated-3d4he-2p4m}).
%Forced vortices are generated in the rotation problem of a single-phase flow in classical fluid mechanics. 
A single forced vortex that spans the container is generated in the rotation problem of a single-phase flow in classical fluid mechanics.
However, it can be confirmed that the behavior of the fluid is similar to that of a quantum fluid. In other words, we observed that the low-density and viscous components coalesced to form multiple spinning vortices. By contrast, in this simulation, the vortices repeat arbitrary coupling and separation while continuously changing. This is characteristic of classical fluids. It is reasonable that the dynamics of the vortex should be similar to those of classical flow because it does not currently contain artificial rules for vortex recombination, as observed in quantum fluids. In the future, if, for example, K. W. Schwarz's rule~\cite{PhysRevB.31.5782, PhysRevB.38.2398} is introduced, it will be possible to get closer to the second image of the quantum vortex. Snapshots from the beginning of the simulation to 0.79 s in physical time in the high-resolution case with approximately 19.6 million particles are shown in Fig.~\ref{fig:Figure-SimHighReso}.
A video of the simulation is presented in Fig.~\ref{fig:Figure-SimHighReso} is provided as integral multimedia in Fig.~\ref{fig:Figure-Video-HighResSim-0160}.
A preprint version of the video is available at (\url{https://www.satoritsuzuki.org/video-horizontally-rotated-3d4he-19m}).
Low-density and viscous components agglomerate to form multiple spinning vortices, which is common in the low-resolution case. This is also the same as in the low-resolution case in that the vortices experience arbitrary connection and separation with continuous change. However, unlike in the low-resolution case, the individual vortices are thinner and the number of vortices is larger, even though the computational conditions are the same.

\section{Discussion}\label{sec:Discuss}
A description of the dynamics of liquid helium-4 at cryogenic temperatures based on classical fluid dynamics yields Eqs.~(\ref{eq:goveqsuper:mut}) and (\ref{eq:goveqnormal:mut}), which are similar to the conventional two-fluid model derived by combining the nonlinear $\rm Schr\ddot{o}dinger$ equation for bosons, that is, the Gross--Pitaevskii equation, with the thermodynamic relation Gibbs--Duhem equation. In addition, our two-fluid model includes a conservation term for angular momentum related to the rotation of classical fluid particles. The fact that we could reproduce multiple spinning vortices, such as quantum vortices, using a method based on classical fluid mechanics would be groundbreaking because multiple spinning vortices and their lattice phenomena in liquid helium-4 were thought to originate purely from quantum mechanics. 
The simulation results showed that the individual vortices were thinner and that there were more vortices in the high-resolution simulation than in the low-resolution simulation. This is reasonable because if we focus on the cross section at z = 0, the number of vortices observed on the cross section should be a few hundred according to Feynman's law in two-dimensional cases~\cite{doi:10.1063/5.0060605}. In addition, the vortex centers of quantum vortices are essentially of the order of angstroms (although quantum vortices are thought to interact even at large distances). Therefore, the results of the high-resolution simulation were closer to the theoretical solution.

As mentioned previously, the usual two-fluid model and our fluid model have the same form of equations, but their interpretations differ. Usually, the two components–superfluid and normal flow–overlap, and the interaction between them is interpreted as occurring only through the mutual friction term. In other words, these two components coexist in a single region. This is a quantum mechanical picture. However, our model considers liquid helium as an inviscid fluid in a continuum, with a viscous component remaining at finite temperatures. The two components interact as a multiphase flow under certain conditions; however, the nature of a multiparticle system appears locally. 
%In this case, the temperature gradient along the horizontal direction was calculated. 
In this case, the temperature gradient along the horizontal direction was considered. 
In addition, a local interaction force proportional to the velocity difference, which is often used in particle-fluid coupling, acts between the two-fluid particles (we also call this the mutual friction force because it has the same form as the quantum mechanical mutual friction force). Thus, multiphase flow and particle-fluid coupling are often modeled according to classical fluid mechanics. 
% Thus, because the fluid particles are only fragments of continuum mechanics, they do not overlap and exhibit an excluded volume effect on each other. 
Because the fluid particles are only fragments of continuum mechanics, they do not overlap and exhibit an excluded volume effect on each other. 
In summary, unlike the typical two-fluid model, our two-fluid model, which is based on classical fluid mechanics, is a one-fluid system. In this sense, it makes sense that high-resolution simulations approach the quantum mechanical picture more closely than low-resolution simulations. In multiparticle simulations, the computational resolution is determined by the diameter of the fluid particles. As the computation increases in resolution, the fluid particles decrease in size. Simultaneously, the excluded volume effect of the fluid particles also decreased; therefore, the particles were more likely to slip through and overlap with each other. Consequently, our method describes a one-fluid system at a low resolution, in which the two components are separated to describe a single fluid. However, as the resolution increases, the system is expected to become a two-fluid system. In other words, we can expect our two-fluid model to be smoothly connected and equal to the two-fluid model. That is, in this problem, the convergence of the calculations coincides with the convergence of the physical phenomena.

In our model, the fluid particles were subjected to pressure gradient forces, temperature gradient forces, shear viscosity, rotational viscosity, and mutual friction forces. The shear and rotational viscous forces act only between the components of the viscous fluid. In a typical case, the distribution of these forces acting on the fluid particles was approximately 72.9 \% pressure gradient force, 8.28 \% temperature gradient force, 18.07 \% shear viscous force, and approximately 0.71 \% rotational viscous force. The mutual friction force was approximately 0.01 \%--0.04 \% that of the rest. Although the rotational viscous force is less than 1 \% of the force applied to the fluid particles, it is understandable that it causes an independently spinning vortex. The mutual frictional force is negligible but may act as a disturbance, and which force induces which physical phenomena should be investigated in future studies.

\begin{figure}[t]
\vspace{-41.0cm}
\hspace{28.0cm}
%\begin{center}
\centerline{\includegraphics[width=4.3\textwidth, clip, bb= 0 0 3456 2470]{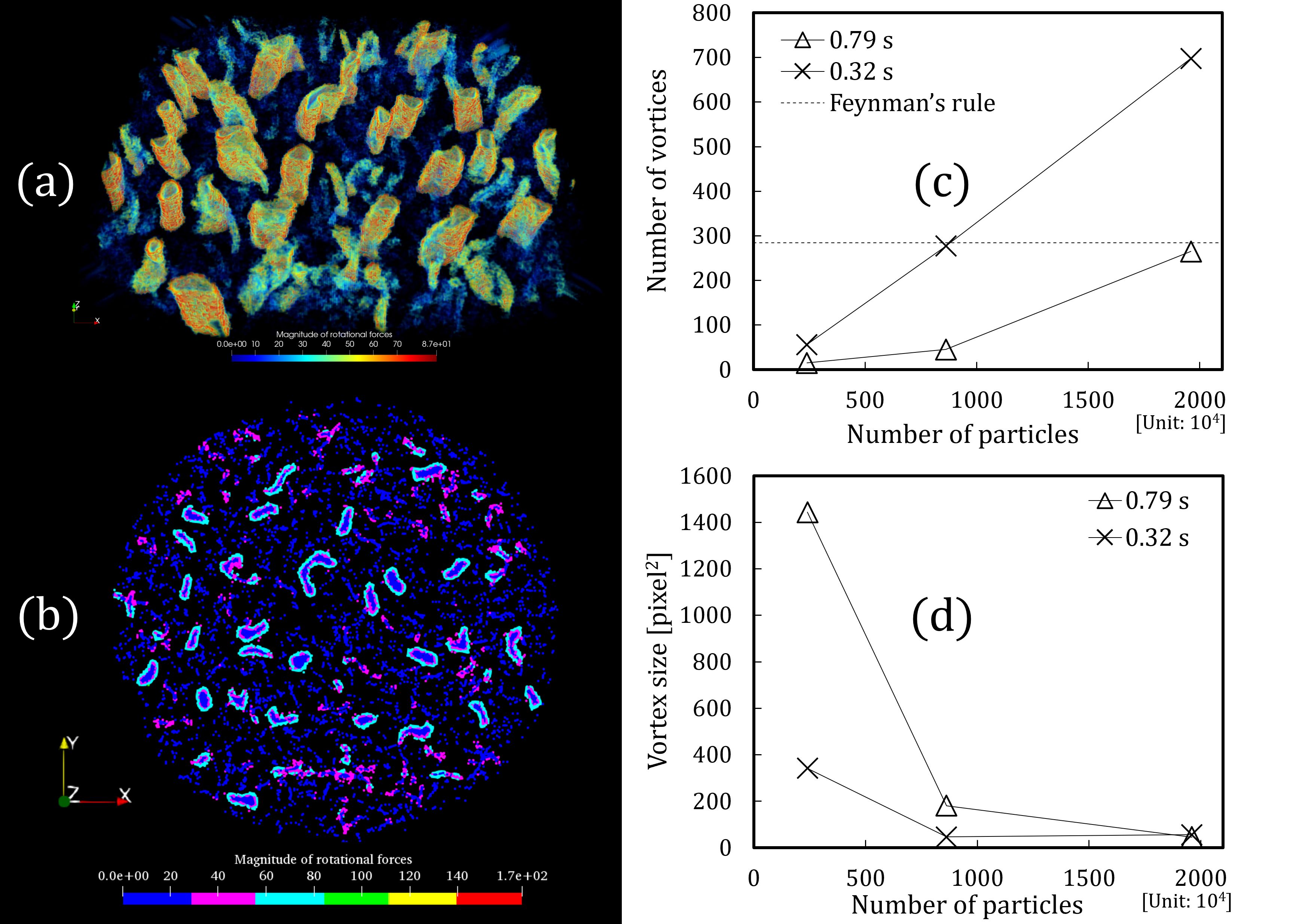}}
\caption{\HL{Schematic explanations of the image analysis to count the number of vortices: (a) an example of the volume rendering, showing the intensity of the rotational force at time 0.32 s in the lower resolution, which corresponds to Fig.~\ref{fig:Figure-SimLowReso}(c), and (b) the intensity distribution of the rotational force at $z = 0$ plane, which corresponds to the scalar distribution shown in (a). (c) the correlation between the number of observed vortices and the number of particles, and (d) the relationship between the average vortex size and the number of particles, at 0.32 and 0.79 s.}}
%\end{center}
\label{fig:Figure-NumVortices}
\end{figure}

\HL{Several experimental studies~\cite{PhysRev.60.356, Yarmchuk1982} on rotated cryogenic liquid helium-4 have reported peculiar phenomena compared to those observed in rotation problems in classical hydrodynamics. Specifically, multiple quantum vortices emerge and spin around their axes parallel to the vertical center axis of the cylindrical vessel, eventually forming a regular lattice. The lattices rotate around the cylindrical axis while maintaining a constant relative position, which is called rigid-body rotation. Of particular interest is the emergence of multiple independent spinning nondissipative vortices; this phenomenon in a 3D problem was numerically reproduced for the first time in this study. To clarify the implications of the findings of this study and for future tasks, we investigated the dependence of the number and average size of vortices on the numerical resolution, that is, the number of fluid particles, at two different elapsed times. The number of vortices was measured in the following manner. Because vortices exhibit vertical structures, we measured the number of vortices that crossed the $z = 0$ plane. We identified and labeled the vortices from the intensity distribution of the rotational forces exerted by the respective vortices on the $z = 0$ plane using the {\tt ImageJ} software~\cite{10030139275, abramoff2004image, Schneider2012}. Figures~\ref{fig:Figure-NumVortices}(a) and (b) present this process; (a) is an example snapshot of the volume rendering of the simulation, showing the intensity of the rotational force at time of 0.32 s in the lower resolution simulation with 2.4 million particles; this subfigure corresponds to Fig.~\ref{fig:Figure-SimLowReso}(c); (b) shows the scalar distribution of the intensity of the rotational force in the $z = 0$ plane in (a). 	
Figures~\ref{fig:Figure-NumVortices}(a) and (b) clearly show that the edges of each vortex can be identified because the intensity of the rotational force is maximized at the edges of the vortices. Finally, Fig.~\ref{fig:Figure-NumVortices}(c) shows the correlation between the numbers of observed vortices and particles, and Fig.~\ref{fig:Figure-NumVortices}(d) shows the relationship between the average vortex size and number of particles at 0.32 and 0.79 s.

The significant observations are summarized as follows. First, vortices tend to move freely or separately, absorbing neighboring smaller vortices and changing into a smaller number of larger vortices; the total number of vortices decreases with time. In brief, local interactions among vortices are classical. In practice, fewer vortices were observed, as shown in Fig.~\ref{fig:Figure-NumVortices}(c) for the triangular symbol (at 0.79 s) compared to that for the cross symbol (at 0.32 s). However, it should be emphasized that independent vortices rarely disappear unlike in ordinary classical fluids. In summary, multiple spinning, independent, and non-dissipative vortices emerged for horizontally rotated 3D cryogenic helium-4 at a lower resolution; however, local interactions among the vortices were classical, except for their non-dissipative nature.

However, this phenomenon was observed less frequently as the resolution increased. In Fig.~\ref{fig:Figure-NumVortices}(d), as the resolution increases from low (2.4 million particles) to medium (8.62 million particles) to high (19.6 million particles), the average size of the vortices decreases and no significant difference in size is observed between two different cases of elapsed time. Specifically, the average size of the vortices can be considered to reach a steady state over time.
First, considering that the proposed method does not introduce constraints on the topology changes in the vortices, the interactions between the vortices inevitably exhibit classical fluidity. Nevertheless, as the resolution increases, the vortices become smaller.
This can be described as follows. In our simulations, the diameter of the fluid particles decreased as the resolution increased, but the value of the parameter $C_{\omega}$, which determines the magnitude of the angular velocity, was set to be inversely proportional to the size of the particle such that the rotational force at the edge of the individual fluid particles, expressed as their product as $r |\vec{\omega_{0}}|$, remained the same. This ensures that the torque (rotational force) on each fluid particle is constant regardless of the resolution. Consequently, although the number of particles required to form a vortex remained constant, the average size of the vortices decreased because the size of the constituent particles decreased. Then, the smaller sized vortices with the same magnitude of rotational force are distributed over the entire domain. Because the average size of the vortices decreases, the frequency of vortex-vortex interactions decreases; a higher number of multiple rotating vortices with small sizes can exist compared to the low-resolution case. 

As we will see later, the average size of the vortices at 19.6 million particles is comparable to that of the average vortex-vortex distance derived from Feynman's rule in quantum hydrodynamics; from a local approximation point of view, it can be considered approximately equivalent to the velocity field generated by each quantum vortex. From a phenomenological perspective, the fact that a number of multiple independent vortices can exist with smaller sizes comparable to the velocity field of quantum vortices as the resolution increases suggests that vortices become more quantum mechanical as the resolution increases.
In other words, our two-fluid model might directly describe the microscopic behavior as the resolution increases. There are two possible explanations for this observation. The first possibility is that this claim is incorrect, and that artificial control of the vortex topology is still required. In quantum mechanics, circulation is quantized. Therefore, we may need to introduce a constraint such as the Schwarz equation to maintain the circulation or each vortex constant, leading to a change in the system into a quantum fluid system. The second possibility is that such a constraint can be introduced automatically. According to the Young-Laplace equation, the surface tension of a classical fluid is proportional to its curvature. As the vortex becomes smaller, its surface tension increases, which may solidify the vortex and act as a type of topological control. To validate this, simulations with the same physical conditions and at higher resolutions must be performed.

Based on the results of this study, the vortices have more opportunities to interact with each other over longer periods. Consequently, the vortices gradually integrate and decrease to the same degree as the ideal number of vortices calculated by Feynman's rule, which is equal to 284.7 in the corresponding 2D case (refer to Ref.~\cite{doi:10.1063/5.0060605} for specific calculations) with 19.6 million particles in Fig.~\ref{fig:Figure-NumVortices}(c). However, even in this case, the number of vortices decreased with time. Thus, the system did not reach a steady state with respect to the number of vortices in the time direction. Among the two possibilities, in the former, a comparison with the theoretical value of Feynman's rule must be discussed in detail after the corresponding 3D physical model of circulation control is introduced, and the number of vortices becomes stable in the time direction. By contrast, in the latter, higher resolution simulations are required. In Fig.~\ref{fig:Figure-NumVortices}(c), the curve with the triangular symbol is exponential relative to the computational resolution, whereas the curve with the cross symbol is linear with respect to the computational resolution. Therefore, we can expect to reach a stationary state in the time direction when using tens to 100 million particles for the entire domain without introducing such a topology control model. 
In the former case, developing a 3D vortex dynamics model based on classical fluid mechanics as an alternative to Schwarz's rule~\cite{PhysRevB.31.5782, PhysRevB.38.2398} is another approach. Our results for the 2D problem have been presented in previous studies~\cite{doi:10.1063/5.0060605, doi:10.1063/5.0122247}, where we proposed a model that balances the Magnus force generated by each rotating vortex with the repulsive force as a vortex-vortex interaction, which prevents the vortices from moving closer than necessary. Specifically, by calculating the vortex circulation value and introducing a numerical model in which the vortex dynamics model becomes effective when the circulation exceeds a certain criterion, we successfully controlled the equal arrangement and circulation of vortices~\cite{doi:10.1063/5.0060605, doi:10.1063/5.0122247}. Hence, in principle, it is possible to develop similar dynamic vortex models for 3D problems. However, several issues need to be resolved, such as the physical correspondence between the classical vortex dynamics model and Schwarz's rule~\cite{PhysRevB.31.5782, PhysRevB.38.2398}, including whether they are in one-to-one correspondence. Moreover,  in future research, our 2D vortex dynamics model should be extended to changes in 3D topology. 

For 19.6 million particles, 500 fluid particles were arranged in each horizontal direction, as mentioned previously. Because the diameter of the disk was 0.2 cm (${\rm = 2.0 \times 10^{-3} m}$), the diameter of a particle is ${\rm 4 \times 10^{-6} m}$ or 4 micrometers. If we assume 10 particles per vortex in each horizontal direction, the size of a single vortex would be approximately 40 micrometers. In contrast, the scale of a vortex core is on the order of angstroms. However, because the effective range of a vortex is determined by the Biot--Savart law, which describes the velocity field around the vortex, the area of influence of the vortex is considered to be its effective size. In our case, the average vortex distance was calculated using Feynman's rule, which yielded approximate ${\rm 10^{-4} m}$~\cite{doi:10.1063/5.0060605}. The corresponding number of vortices is shown by the dotted line in Fig.~\ref{fig:Figure-NumVortices}(c). It is reasonable to understand that the vortex observed does not correspond to an actual quantum vortex core, but to the effective velocity field created by the quantum vortex. 

In summary, the next task is to control the circulation of vortices by ensuring an appropriate vortex–vortex interaction to stabilize the number of vortices in the time direction. As desctibed above, this may be achieved by introducing an artificial circulation control model by referring to Schwarz's rule, or by simply increasing the resolution, expecting that an increase in the strength of the surface tension would solidify the vortex. We regard the latter case as promising for the following reasons. In Fig.~\ref{fig:Figure-NumVortices}(c), the curve with the triangular symbol is exponential relative to the computational resolution, whereas that with the cross symbol is linear with respect to the computational resolution. Extrapolating these results enables us to expect that the vortices may reach an approximately stationary state in the time direction when tens to 100 million particles are used in the entire domain. Another approach is to develop a 3D vortex dynamics model based on classical hydrodynamics, as an alternative to Schwarz's rule. We achieved some success in the 2D calculations using models representing this effect in the Magnus force and vortex repulsion models~\cite{doi:10.1063/5.0060605}, as described above. However, several issues need to be resolved, such as the physical correspondence between the classical vortex dynamics model and Schwarz's rule and how to extend our 2D vortex dynamics model to changes in 3D topology. Nevertheless, this approach should be examined after checking whether the second approach is valid.

From an engineering perspective, our goal is not to propose a first-principles model, but to enable the simulation of practical bulk liquid helium; we only have to capture the dynamic interactions among vortices accurately enough to reproduce macroscopic quantum effects. In practice, a direct comparison between the internal structure within the healing length of the quantum vortex and the vortex in our method requires at least several hundred times more particles in each direction, even when using a coarse-grained approach. However, this was beyond the scope of this study. At present, this is the first study to numerically reproduce the dynamics of rotated cryogenic 3D liquid helium-4 in bulk form using a classical hydrodynamic approach. Accordingly, it is scientifically significant that we observed multiple spinning nondissipative vortices in the 3D problem of horizontally rotated cryogenic liquid helium-4.
}

\section{Conclusion}
In this paper, we report a 3D simulation of the rotating liquid helium-4 using a two-fluid model with spin-angular momentum conservation using SPH. The conventional two-fluid model was derived by combining the nonlinear Schrodinger equation for bosons with the thermodynamic Gibbs--Duhem equation. In short, the usual two-fluid model is obtained by making macro-corrections to the microscopic relational equations. In contrast, our model is derived from the particle approximation of an inviscid fluid with residual viscosity in part. Despite the fully classical mechanical picture, the resulting system equations were consistent with those of the conventional two-fluid model. Specifically, we consider the bulk liquid helium-4 to be an inviscid fluid, assuming that the viscous fluid component remains at finite temperatures. As the temperature decreased, the amount of the viscous fluid component decreased, ultimately becoming a fully inviscid fluid at absolute zero. Weak compressibility is assumed to express the volume change because some helium atoms do not render fluid owing to BECs or change states owing to local thermal excitation. Using explicit SPH (smoothed-particle hydrodynamics), one can solve the governing equations for an incompressible fluid, simultaneously reproducing density fluctuations and describing the fluid in a many-particle system. 
We assume the following fluid-particle duality: a hydrodynamic interfacial tension between the inviscid and viscous components or a local interaction force between two types of fluid particles. The former can be induced in the horizontal direction when non-negligible non-uniformity of the particles occurs during forced 2D rotation, and the latter is non-negligible when the former is negligible. We performed a large-scale simulation of 3D liquid helium forced to rotate horizontally using 32 GPUs. Compared with the low-resolution calculation using 2.4 million particles, the high-resolution calculation using 19.6 million particles showed spinning vortices close to those of the theoretical solution. In the future, the introduction of quantum mechanical recombination among vortices will capture the behavior of quantum fluids. We obtained a promising venue to establish a practical simulation method for bulk liquid helium-4.

\section*{Acknowledgment}
This study was supported by JSPS KAKENHI Grant Number 22K14177 and JST PRESTO, Grant Number JPMJPR23O7.
The authors thank Editage (www.editage.jp) for the English language editing.
The author would also like to express his gratitude to his family for their moral support and encouragement.

\appendix
\section*{Appendix}
\begin{figure}[t]
\vspace{-34.4cm}
\hspace{29.3cm}
%\begin{center}
\centerline{\includegraphics[width=4.56\textwidth, clip, bb= 0 0 4419 2508]{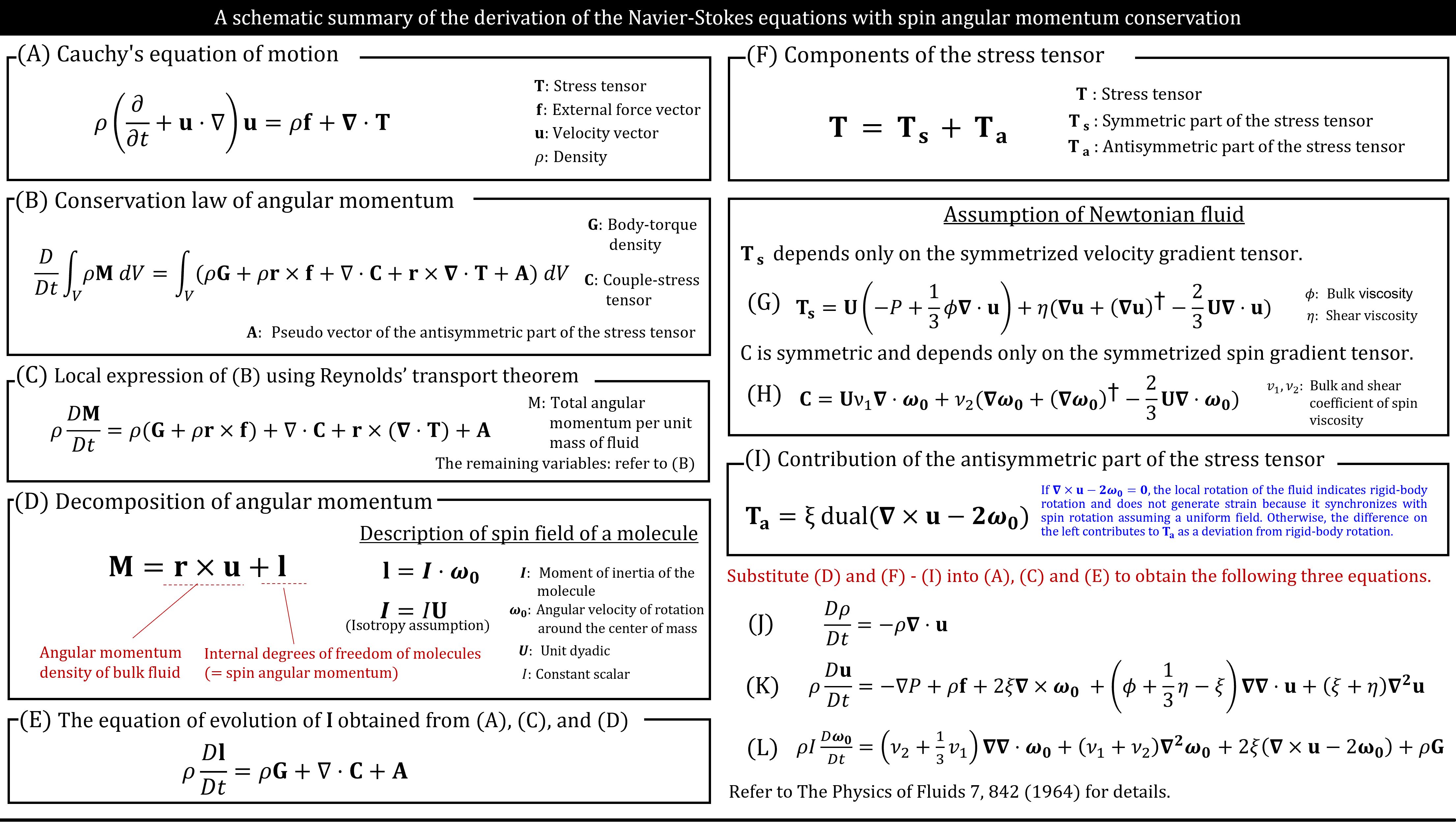}}
\caption{\HL{A schematic list of the equations for the derivation of the Navier--Stokes equations with spin angular momentum conservation in the literature~\cite{doi:10.1063/1.1711295}: (A) Cauchy's equation of motion, (B) the conservation law of angular momentum, (C) the local expression of (B) using Reynolds' transport theorem, (D) a decomposition of the angular momentum, (E) the evolution equation of $\vec{I}$ obtained from (A), (C) and (D), (F) two components of the stress tensor, and (D), (F) two components of the stress tensor, (G) assumption of Newtonian fluid for the stress tensor and (H) for the coupled stress tensor, (I) contribution of the antisymmetric part of the stress tensor, and the resulting equations in (J), (K), and (L).}}
%\end{center}
\label{fig:Figure-NSeqWithSpinConsv}
\end{figure}

\begin{figure}[t]
\vspace{-34.4cm}
\hspace{29.3cm}
%\begin{center}
\centerline{\includegraphics[width=4.56\textwidth, clip, bb= 0 0 4416 2476]{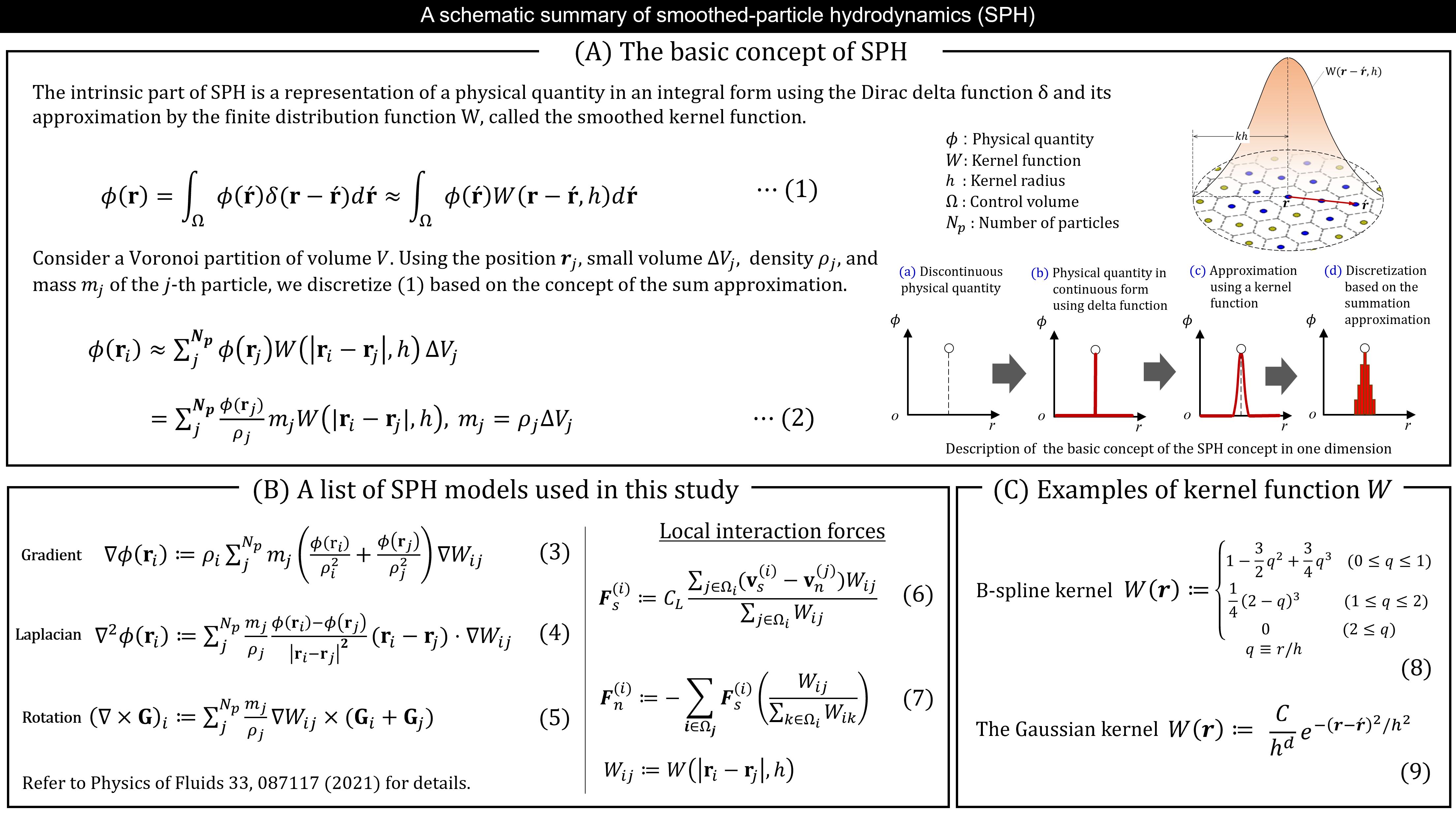}}
\caption{\HL{A schematic summary of SPH (smoothed-particle hydrodynamics): (A) the basic concept of SPH, (B) a list of SPH models used in this study, and (C) examples of kernel functions.}}
%\end{center}
\label{fig:Figure-SchemOfSPH}
\end{figure}

\HL{To facilitate understanding, we provide a schematic list of the equations for the derivation of the Navier--Stokes equations with spin angular momentum conservation in Fig.~\ref{fig:Figure-NSeqWithSpinConsv}. We also provide a schematic summary of SPH in Fig.~\ref{fig:Figure-SchemOfSPH}.}

%\section*{References}
%\bibliographystyle{elsarticle-num} 
\bibliographystyle{h-physrev3}
\bibliography{reference}

\begin{thebibliography}{10}

\bibitem{Arzhavitin2016}
V.~M. Arzhavitin,
\newblock Diffusion mechanism of internal friction in a niobium--titanium
  alloy, Russian Metallurgy (Metally) {\bf 2016}, 431 (2016).

\bibitem{BANNO2023100047}
N.~Banno,
\newblock Low-temperature superconductors: Nb3sn, nb3al, and nbti,
  Superconductivity {\bf 6}, 100047 (2023).

\bibitem{10.1063/10.0020164}
A.~T. Jones, R.~B.~E. Down, C.~R. Lawson, D.~Keeping, and O.~Kirichek,
\newblock {Carbon footprint of helium recovery systems}, Low Temperature
  Physics {\bf 49}, 967 (2023),
  https://pubs.aip.org/aip/ltp/article-pdf/49/8/967/18070432/967\_1\_10.0020164.pdf.

\bibitem{WilsonLloyd1950}
L.~H. Wilson,
\newblock {\em The Viscosity of Gases},
\newblock PhD thesis, 1950.

\bibitem{Immanuil-L-Fabelinskii_1997}
I.~L. Fabelinski^^c4^^ad,
\newblock Macroscopic and molecular shear viscosity, Physics-Uspekhi {\bf 40},
  689 (1997).

\bibitem{Tsuzuki_2021}
S.~Tsuzuki,
\newblock Particle approximation of the two-fluid model for superfluid 4he
  using smoothed particle hydrodynamics, Journal of Physics Communications {\bf
  5}, 035001 (2021).

\bibitem{doi:10.1063/5.0060605}
S.~Tsuzuki,
\newblock Reproduction of vortex lattices in the simulations of rotating liquid
  helium-4 by numerically solving the two-fluid model using smoothed-particle
  hydrodynamics incorporating vortex dynamics, Physics of Fluids {\bf 33},
  087117 (2021), https://doi.org/10.1063/5.0060605.

\bibitem{doi:10.1063/5.0122247}
S.~Tsuzuki,
\newblock Theoretical framework bridging classical and quantum mechanics for
  the dynamics of cryogenic liquid helium-4 using smoothed-particle
  hydrodynamics, Physics of Fluids {\bf 34}, 127116 (2022),
  https://doi.org/10.1063/5.0122247.

\bibitem{doi:10.1063/1.1711295}
D.~W. Condiff and J.~S. Dahler,
\newblock Fluid mechanical aspects of antisymmetric stress, The Physics of
  Fluids {\bf 7}, 842 (1964).

\bibitem{MULLER2015301}
K.~M{\"{u}}ller, D.~A. Fedosov, and G.~Gompper,
\newblock Smoothed dissipative particle dynamics with angular momentum
  conservation, Journal of Computational Physics {\bf 281}, 301  (2015).

\bibitem{gingold1977smoothed}
R.~A. Gingold and J.~J. Monaghan,
\newblock Smoothed particle hydrodynamics: theory and application to
  non-spherical stars, Monthly notices of the royal astronomical society {\bf
  181}, 375 (1977).

\bibitem{monaghan1992smoothed}
J.~J. Monaghan,
\newblock Smoothed particle hydrodynamics, Annual review of astronomy and
  astrophysics {\bf 30}, 543 (1992).

\bibitem{PhysRevB.31.5782}
K.~W. Schwarz,
\newblock Three-dimensional vortex dynamics in superfluid $^{4}\mathrm{He}$:
  Line-line and line-boundary interactions, Phys. Rev. B {\bf 31}, 5782 (1985).

\bibitem{PhysRevB.38.2398}
K.~W. Schwarz,
\newblock Three-dimensional vortex dynamics in superfluid $^{4}\mathrm{He}$:
  Homogeneous superfluid turbulence, Phys. Rev. B {\bf 38}, 2398 (1988).

\bibitem{Darve2011}
C.~M.-T. Darve,
\newblock {\em Phenomenological and Numerical Studies of Helium II Dynamics in
  the Two-Fluid Model},
\newblock PhD thesis, United States -- Illinois, 2011,
\newblock Ph.D.

\bibitem{TISZA1938}
L.~TISZA,
\newblock Transport phenomena in helium ii, Nature {\bf 141}, 913 (1938).

\bibitem{PhysRev.60.356}
L.~Landau,
\newblock Theory of the superfluidity of helium ii, Phys. Rev. {\bf 60}, 356
  (1941).

\bibitem{GORTER1949285}
C.~Gorter and J.~Mellink,
\newblock On the irreversible processes in liquid helium ii, Physica {\bf 15},
  285  (1949).

\bibitem{PhysRevLett.49.279}
V.~F. Sears, E.~C. Svensson, P.~Martel, and A.~D.~B. Woods,
\newblock Neutron-scattering determination of the momentum distribution and the
  condensate fraction in liquid $^{4}\mathrm{He}$, Phys. Rev. Lett. {\bf 49},
  279 (1982).

\bibitem{Monaghan_2005}
J.~J. Monaghan,
\newblock Smoothed particle hydrodynamics, Reports on Progress in Physics {\bf
  68}, 1703 (2005).

\bibitem{becker2007weakly}
M.~Becker and M.~Teschner,
\newblock Weakly compressible sph for free surface flows,
\newblock in {\em Proceedings of the Eurographics symposium on Computer
  animation}, pp. 209--217, Eurographics Association, 2007.

\bibitem{MONAGHAN1994399}
J.~Monaghan,
\newblock Simulating free surface flows with sph, Journal of Computational
  Physics {\bf 110}, 399  (1994).

\bibitem{NOMERITAE2016156}
Nomeritae, E.~Daly, S.~Grimaldi, and H.~H. Bui,
\newblock Explicit incompressible sph algorithm for free-surface flow
  modelling: A comparison with weakly compressible schemes, Advances in Water
  Resources {\bf 97}, 156  (2016).

\bibitem{Idowu2001}
O.~C. Idowu, D.~Kivotides, C.~F. Barenghi, and D.~C. Samuels,
\newblock {\em Numerical Methods for Coupled Normal-Fluid and Superfluid Flows
  in Helium II} (Springer Berlin Heidelberg, Berlin, Heidelberg, 2001), pp.
  162--176.

\bibitem{PhysRevLett.120.155301}
S.~Yui, M.~Tsubota, and H.~Kobayashi,
\newblock Three-dimensional coupled dynamics of the two-fluid model in
  superfluid $^{4}\mathrm{He}$: Deformed velocity profile of normal fluid in
  thermal counterflow, Phys. Rev. Lett. {\bf 120}, 155301 (2018).

\bibitem{doi:10.1063/1.5091567}
C.~L. Horner and R.~A. Van~Gorder,
\newblock Dynamics of quantized vortex filaments under a local induction
  approximation with second-order correction, Physics of Fluids {\bf 31},
  065103 (2019), https://doi.org/10.1063/1.5091567.

\bibitem{PhysRevLett.124.155301}
S.~Yui, H.~Kobayashi, M.~Tsubota, and W.~Guo,
\newblock Fully coupled two-fluid dynamics in superfluid $^{4}\mathrm{He}$:
  Anomalous anisotropic velocity fluctuations in counterflow, Phys. Rev. Lett.
  {\bf 124}, 155301 (2020).

\bibitem{doi:10.1073/pnas.1608074113}
L.~Lanotte {\em et~al.},
\newblock Red cells' dynamic morphologies govern blood shear thinning under
  microcirculatory flow conditions, Proceedings of the National Academy of
  Sciences {\bf 113}, 13289 (2016).

\bibitem{NEMIROVSKII201385}
S.~K. Nemirovskii,
\newblock Quantum turbulence: Theoretical and numerical problems, Physics
  Reports {\bf 524}, 85  (2013),
\newblock Quantum Turbulence: Theoretical and Numerical Problems.

\bibitem{Adamenko_2008}
I.~N. Adamenko, K.~E. Nemchenko, I.~V. Tanatarov, and A.~F.~G. Wyatt,
\newblock Pressure of thermal excitations in superfluid helium, Journal of
  Physics: Condensed Matter {\bf 20}, 245103 (2008).

\bibitem{schmitt2015introduction}
A.~Schmitt,
\newblock Introduction to superfluidity, Lect. Notes Phys {\bf 888} (2015).

\bibitem{bennemann2013novel}
K.-H. Bennemann and J.~B. Ketterson,
\newblock {\em Novel superfluids}volume~1 (OUP Oxford, 2013).

\bibitem{hammond2000elements}
C.~Hammond,
\newblock The elements, Handbook of chemistry and physics {\bf 81} (2000).

\bibitem{10.5555/2388996.2389003}
T.~Ishiyama, K.~Nitadori, and J.~Makino,
\newblock 4.45 pflops astrophysical n-body simulation on k computer: the
  gravitational trillion-body problem,
\newblock in {\em Proceedings of the International Conference on High
  Performance Computing, Networking, Storage and Analysis}, SC '12, Washington,
  DC, USA, 2012, IEEE Computer Society Press.

\bibitem{Sugimoto1990}
D.~Sugimoto {\em et~al.},
\newblock A special-purpose computer for gravitational many-body problems,
  Nature {\bf 345}, 33 (1990).

\bibitem{https://doi.org/10.1002/jcc.20289}
J.~C. Phillips {\em et~al.},
\newblock Scalable molecular dynamics with namd, Journal of Computational
  Chemistry {\bf 26}, 1781 (2005).

\bibitem{valiev2010nwchem}
M.~Valiev {\em et~al.},
\newblock Nwchem: A comprehensive and scalable open-source solution for large
  scale molecular simulations, Computer Physics Communications {\bf 181}, 1477
  (2010).

\bibitem{Heinecke2015}
A.~Heinecke, W.~Eckhardt, M.~Horsch, and H.-J. Bungartz,
\newblock {\em Parallelization of MD Algorithms and Load Balancing} (Springer
  International Publishing, Cham, 2015), pp. 31--44.

\bibitem{10.1007/978-3-031-39698-4_38}
Z.~Zhang {\em et~al.},
\newblock Swsph: A massively parallel sph implementation
  for^^c2^^a0hundred-billion-particle simulation on^^c2^^a0new sunway
  supercomputer,
\newblock in {\em Euro-Par 2023: Parallel Processing}, edited by J.~Cano, M.~D.
  Dikaiakos, G.~A. Papadopoulos, M.~Peric{\`a}s, and R.~Sakellariou, pp.
  564--577, Cham, 2023, Springer Nature Switzerland.

\bibitem{OGER20161}
G.~Oger {\em et~al.},
\newblock On distributed memory mpi-based parallelization of sph codes in
  massive hpc context, Computer Physics Communications {\bf 200}, 1 (2016).

\bibitem{CHEN1996195}
J.-S. Chen, C.~Pan, C.-T. Wu, and W.~K. Liu,
\newblock Reproducing kernel particle methods for large deformation analysis of
  non-linear structures, Computer Methods in Applied Mechanics and Engineering
  {\bf 139}, 195 (1996).

\bibitem{Nayroles1992}
B.~Nayroles, G.~Touzot, and P.~Villon,
\newblock Generalizing the finite element method: Diffuse approximation and
  diffuse elements, Computational Mechanics {\bf 10}, 307 (1992).

\bibitem{https://doi.org/10.1002/nme.1620370205}
T.~Belytschko, Y.~Y. Lu, and L.~Gu,
\newblock Element-free galerkin methods, International Journal for Numerical
  Methods in Engineering {\bf 37}, 229 (1994).

\bibitem{Lin2020}
Z.~Lin, D.~Wang, D.~Qi, and L.~Deng,
\newblock A petrov-galerkin finite element-meshfree formulation for
  multi-dimensional fractional diffusion equations, Computational Mechanics
  {\bf 66}, 323 (2020).

\bibitem{MikioSakai20202020017}
M.~Sakai, Y.~Mori, X.~Sun, and K.~Takabatake,
\newblock Recent progress on mesh-free particle methods for simulations of
  multi-phase flows: A review, KONA Powder and Particle Journal {\bf 37}, 132
  (2020).

\bibitem{doi:10.1680/geot.1979.29.1.47}
P.~A. Cundall and O.~D.~L. Strack,
\newblock A discrete numerical model for granular assemblies,
  G^^c3^^a9otechnique {\bf 29}, 47 (1979).

\bibitem{PARK2021104008}
E.~H. Park, V.~Kindratenko, and Y.~M. Hashash,
\newblock Shared memory parallelization for high-fidelity large-scale 3d
  polyhedral particle simulations, Computers and Geotechnics {\bf 137}, 104008
  (2021).

\bibitem{Yan2018}
B.~Yan and R.~A. Regueiro,
\newblock A comprehensive study of mpi parallelism in three-dimensional
  discrete element method (dem) simulation of complex-shaped granular
  particles, Computational Particle Mechanics {\bf 5}, 553 (2018).

\bibitem{GAN2020258}
J.~Gan, T.~Evans, and A.~Yu,
\newblock Application of gpu-dem simulation on large-scale granular handling
  and processing in ironmaking related industries, Powder Technology {\bf 361},
  258 (2020).

\bibitem{GREST1989269}
G.~S. Grest, B.~D^^c3^^bcnweg, and K.~Kremer,
\newblock Vectorized link cell fortran code for molecular dynamics simulations
  for a large number of particles, Computer Physics Communications {\bf 55},
  269 (1989).

\bibitem{10.1016/j.cageo.2012.02.028}
M.~Gomez-Gesteira {\em et~al.},
\newblock Sphysics - development of a free-surface fluid solver - part 2:
  Efficiency and test cases, Comput. Geosci. {\bf 48}, 300^^e2^^80^^93307
  (2012).

\bibitem{Tsuzuki:2016:EDL:3019094.3019095}
S.~Tsuzuki and T.~Aoki,
\newblock Effective dynamic load balance using space-filling curves for
  large-scale sph simulations on gpu-rich supercomputers,
\newblock in {\em Proceedings of the 7th Workshop on Latest Advances in
  Scalable Algorithms for Large-Scale Systems}, ScalA '16, pp. 1--8,
  Piscataway, NJ, USA, 2016, IEEE Press.

\bibitem{doi:10.1177/1094342017738610}
G.~Bell, D.~H. Bailey, J.~Dongarra, A.~H. Karp, and K.~Walsh,
\newblock A look back on 30 years of the gordon bell prize, The International
  Journal of High Performance Computing Applications {\bf 31}, 469 (2017).

\bibitem{4541126}
D.~Luebke,
\newblock Cuda: Scalable parallel programming for high-performance scientific
  computing,
\newblock in {\em 2008 5th IEEE International Symposium on Biomedical Imaging:
  From Nano to Macro}, pp. 836--838, 2008.

\bibitem{mpi41}
{Message Passing Interface Forum},
\newblock {\em {MPI}: A Message-Passing Interface Standard Version 4.1}, 2023.

\bibitem{doi:10.1137/S1064827595287997}
G.~Karypis and V.~Kumar,
\newblock A fast and high quality multilevel scheme for partitioning irregular
  graphs, SIAM Journal on Scientific Computing {\bf 20}, 359 (1998).

\bibitem{Karypis2011}
G.~Karypis,
\newblock {\em METIS and ParMETIS} (Springer US, Boston, MA, 2011), pp.
  1117--1124.

\bibitem{ZoltanShortTutorial}
K.~Devine, E.~Boman, L.~Riesen, U.~Catalyurek, and C.~Chevalier,
\newblock Getting started with zoltan: A short tutorial,
\newblock in {\em Proc. of 2009 Dagstuhl Seminar on Combinatorial Scientific
  Computing}, 2009,
\newblock Also available as Sandia National Labs Tech Report SAND2009-0578C.

\bibitem{10.1109/5992.988653}
K.~Devine, E.~Boman, R.~Heapby, B.~Hendrickson, and C.~Vaughan,
\newblock Zoltan data management service for parallel dynamic applications,
  Computing in Science and Engg. {\bf 4}, 90^^e2^^80^^9397 (2002).

\bibitem{KDTREE9597006}
R.~Trobec and M.~Depolli,
\newblock A k-d tree based partitioning of computational domains for efficient
  parallel computing,
\newblock in {\em 2021 44th International Convention on Information,
  Communication and Electronic Technology (MIPRO)}, pp. 284--290, 2021.

\bibitem{Dubinski1996APT}
J.~Dubinski,
\newblock A parallel tree code, New Astronomy {\bf 1}, 133 (1996).

\bibitem{ORBLiu2001}
P.~Liu and S.~Bhatt,
\newblock Experiences with parallel n-body simulation, Parallel and Distributed
  Systems, IEEE Transactions on {\bf 11}, 1306  (2001).

\bibitem{5333803}
M.~Reumann {\em et~al.},
\newblock Orthogonal recursive bisection data decomposition for high
  performance computing in cardiac model simulations: Dependence on anatomical
  geometry,
\newblock in {\em 2009 Annual International Conference of the IEEE Engineering
  in Medicine and Biology Society}, pp. 2799--2802, 2009.

\bibitem{SFCDecomp1997-Aluru}
S.~Aluru and F.~E. Sevilgen,
\newblock Parallel domain decomposition and load balancing using space-filling
  curves,
\newblock in {\em Proceedings Fourth International Conference on
  High-Performance Computing}, pp. 230--235, IEEE, 1997.

\bibitem{PhysRev.159.98}
L.~Verlet,
\newblock Computer "experiments" on classical fluids. i. thermodynamical
  properties of lennard-jones molecules, Phys. Rev. {\bf 159}, 98 (1967).

\bibitem{10.5555/76990}
M.~P. Allen and D.~J. Tildesley,
\newblock {\em Computer Simulation of Liquids} (Clarendon Press, USA, 1989).

\bibitem{imoto2019convergence}
Y.~Imoto, S.~Tsuzuki, and D.~Nishiura,
\newblock Convergence study and optimal weight functions of an explicit
  particle method for the incompressible navier--stokes equations,
  Computational Particle Mechanics {\bf 6}, 671 (2019).

\bibitem{SHAKIBAEINIA201213}
A.~Shakibaeinia and Y.-C. Jin,
\newblock Mps mesh-free particle method for multiphase flows, Computer Methods
  in Applied Mechanics and Engineering {\bf 229-232}, 13  (2012).

\bibitem{Vinen2002}
W.~F. Vinen and J.~J. Niemela,
\newblock Quantum turbulence, Journal of Low Temperature Physics {\bf 128}, 167
  (2002).

\bibitem{XU2013101}
X.~Xu, J.~Ouyang, B.~Yang, and Z.~Liu,
\newblock Sph simulations of three-dimensional non-newtonian free surface
  flows, Computer Methods in Applied Mechanics and Engineering {\bf 256}, 101
  (2013).

\bibitem{XU201643}
X.~Xu and X.-L. Deng,
\newblock An improved weakly compressible sph method for simulating free
  surface flows of viscous and viscoelastic fluids, Computer Physics
  Communications {\bf 201}, 43  (2016).

\bibitem{GHIA1982387}
U.~Ghia, K.~Ghia, and C.~Shin,
\newblock High-re solutions for incompressible flow using the navier-stokes
  equations and a multigrid method, Journal of Computational Physics {\bf 48},
  387  (1982).

\bibitem{Yarmchuk1982}
E.~J. Yarmchuk and R.~E. Packard,
\newblock Photographic studies of quantized vortex lines, Journal of Low
  Temperature Physics {\bf 46}, 479 (1982).

\bibitem{10030139275}
W.~RASBAND,
\newblock Imagej, u.s. national institutes of health, bethesda, maryland, usa,
  http://imagej.nih.gov/ij/  (2011).

\bibitem{abramoff2004image}
D.~M.~D. Abr^^c3^^a0moff, D.~P.~J. Magalh^^c3^^a3es, and D.~S.~J. Ram,
\newblock Image processing with imagej, Biophotonics International {\bf 11}, 36
  (2004).

\bibitem{Schneider2012}
C.~A. Schneider, W.~S. Rasband, and K.~W. Eliceiri,
\newblock Nih image to imagej: 25 years of image analysis, Nature Methods {\bf
  9}, 671 (2012).

\end{thebibliography}

\end{document}